\documentclass{article}
\usepackage{ijcai24}

\usepackage{times}
\usepackage{soul}
\usepackage{url}
\usepackage[hidelinks]{hyperref}
\usepackage[utf8]{inputenc}
\usepackage[small]{caption}
\usepackage{graphicx}
\usepackage{amsmath}
\usepackage{amsthm}
\usepackage{booktabs}
\usepackage{algorithm}
\usepackage{algorithmic}
\usepackage[switch]{lineno}
\usepackage{cite}
\usepackage{natbib}
\usepackage{amssymb}
\usepackage{enumitem}
\usepackage{multirow}
\usepackage{mathtools}
\usepackage{xcolor}
\usepackage{setspace}

\newtheorem{proposition}{Proposition}

\title{DGR: A General Graph  Desmoothing Framework for Recommendation via Global and Local Perspectives}

\author{
Leilei Ding$^1$
\and
Dazhong Shen$^2$\and
Chao Wang$^{3,1}$\thanks{Corresponding authors}
\and
Tianfu Wang$^1$\and
Le Zhang$^3$\and
Yanyong Zhang$^1$$^*$
\affiliations
$^1$School of Computer Science and Technology, University of Science and Technology of China, China
$^2$Shanghai AI lab
$^3$Baidu AI lab
$^4$Guangzhou HKUST Fok Ying Tung Research Institute, China\\
\emails
\{tianfuwang,dingleilei\}@mail.ustc.edu.cn, 
dazh.shen@gmail.com,
chadwang2012@gmail.com,
zhangle09@baidu.com,
yanyongz@ustc.edu.cn
}

\begin{document}

\maketitle

\vspace{-3mm}
\begin{abstract}
Graph Convolutional Networks (GCNs) have become pivotal in recommendation systems for learning user and item embeddings by leveraging the user-item interaction graph's node information and topology. However, these models often face the famous over-smoothing issue, leading to indistinct user and item embeddings and reduced personalization. Traditional desmoothing methods in GCN-based systems are model-specific, lacking a universal solution. This paper introduces a novel, model-agnostic approach named \textbf{D}esmoothing Framework for \textbf{G}CN-based \textbf{R}ecommendation Systems (\textbf{DGR}). It effectively addresses over-smoothing on general GCN-based recommendation models by considering both global and local perspectives. Specifically, 
we first introduce vector perturbations during each message passing layer to penalize the tendency of node embeddings approximating overly to be similar with the guidance of the global topological structure. Meanwhile,  we further develop a tailored-design loss term for the readout embeddings to preserve the local collaborative relations between users and their neighboring items. 
In particular, items that exhibit a high correlation with neighboring items are also incorporated to enhance the local topological information.
To validate our approach, we conduct extensive experiments on 5 benchmark datasets based on 5 well-known GCN-based recommendation models, demonstrating the effectiveness and generalization of our proposed framework. Our code is available at GitHub\footnote{\href{https://github.com/me-sonandme/DGR.git}{https://github.com/me-sonandme/DGR.git}}.


\end{abstract}

\vspace{-2mm}
\section{Introduction}
Personalized recommendation is a vital technology in various domains, such as social media, e-commerce, and career recommendation, aiming to recommend items based on users' interests by mining and matching the embeddings of users and candidate items~\cite{wangperson,wang2021variable,travel_recomendation}.
The interaction information between users and items can naturally be represented as a bipartite graph, making Graph Convolutional Networks (GCNs)~\cite{he2020lightgcn} an effective approach to learning user and item embedding vectors. 
Specifically,
conventional GCN-based recommendation models, such as NGCF \cite{ngcf}, LightGCN \cite{he2020lightgcn} and DGCF \cite{wang2020disentangled} exclusively leverage the message passing mechanism to aggregate local information and directly incorporate the signals captured from historical interaction data into the embedding vectors, which have witnessed both the training efficiency and better performance.

Unfortunately, as mentioned in \cite{chen2020revisiting, liu2021interest, chen2022graph, cai2023lightgcl,zhou2023layer, peng2022svd, wang2023see,xia2023graph}, GCN-based recommendation models face the inherent over-smoothing problem, where all user and item embeddings converge toward similarity during the message passing process. It causes a failure in personalized recommendation, resulting in the same recommendation results for every user \cite{zhou2023layer}. 
Moreover, data in recommendation systems are often highly sparse, making it more difficult to model relationships between users and items due to limited direct user-item interaction information. 
Consequently, the existing GCN-based recommendation struggles to accurately capture the differences in the fine-grained relationships among users and items.

Recently, significant efforts have been devoted to addressing the over-smoothing problem in GCN-based recommendation.
For example, many well-known GCN-based recommendation models, such as LightGCN \cite{he2020lightgcn} and NGCF \cite{ngcf}, opt to incorporate outputted embeddings from each layer into the final readout embeddings, preserving node information characteristics and slightly alleviating the over-smoothing problem. 
Meanwhile, there are also some works proposed to adjust the message passing processes~\cite{he2023simplifying,peng2022svd,liu2021interest} or adopt self-supervised learning for maintaining the distinctiveness of user and item embeddings~\cite{xia2023graph,xia2022hypergraph,cai2023lightgcl,sgl,ncl}. 
However, most of these approaches are specifically designed for individual models, leaving a strong need for a general desmoothing framework in GCN-based recommendation systems.

Indeed, to make full use of the sparse information from the user-item interactions, it is essential to take into consideration both the global and local topological perspectives with model-free solutions. 
Specifically, on the one hand, with the mathematical nature of the global topological structure,  the message passing mechanisms in existing GCN models often lead to the rapid convergence of all node embeddings and cause over-something. Therefore, the first challenge lies in modifying the message passing layer to distinguish user item embeddings.
On the other hand, local collaborative relations, which encourage similar embeddings for the neighbors,  are also critical for recommendation performance \cite{wang2023setrank}. It is also challenging to mine those collaborative signals from the local topological structure during desmoothing process.

In this paper, we introduce the \textbf{D}esmoothing Framework for \textbf{G}CN-based \textbf{R}ecommendation Systems (\textbf{DGR}), a novel, universally applicable, and easy-to-use approach to tackle the over-smoothing problem from both global and local perspectives and enhance personalized recommendation in GCN-based recommendation systems. Our approach encompasses two primary modules: \textbf{G}lobal Desmoothing \textbf{M}essage \textbf{P}assing (\textbf{GMP}) and \textbf{L}ocal Node \textbf{E}mbedding \textbf{C}orrection (\textbf{LEC}). The GMP module is designed to counteract the global trend of user and item embeddings approximating overly to the limit state of infinite-layer graph convolutions, i.e., the over-smoothing point. It introduces vector perturbations at each layer to steer embeddings away from the over-smoothing point. Simultaneously, the LEC module focuses on preserving and emphasizing the local collaborative signals, particularly in sparse data scenarios. It works by maintaining local collaborative relationships between users and neighboring items and incorporating highly correlated items to strengthen the local topological information.
Differing from previous methods that targeted improvements in individual GCN-based recommendation systems, our DGR framework is a general, model-free solution, easily adaptable across various GCN-based models. To substantiate our claims, we have rigorously tested DGR with five popular recommendation models across five public benchmark datasets. The results from these extensive experiments, along with several case studies, have convincingly demonstrated the effectiveness and broad applicability of our framework in enhancing personalized recommendations in GCN-based systems.

\vspace{-2mm}
\section{Preliminaries}
In this section, we first introduce the essential background knowledge about GCN-based recommendation systems and the concept of over-smoothing point.

\vspace{-2mm}
\subsection{GCNs in Recommendation Systems}
GCNs are the representative models for complex graph-structured data \cite{tfjou,chen2022entity,chen2022msnea} and have obtained great success in recommendation, where the interaction information between users and items can naturally be represented as a bipartite graph \cite{wu2024afdgcf}.
LightGCN is one of the foundational frameworks for many GCN-based recommendation models \cite{cai2023lightgcl,yu2023xsimgcl,shuai2022review}. It simplifies GCNs by eliminating non-linear feature transformations, retaining only the message passing. Each layer is defined as: 
\begin{equation}
\scalebox{1.0}{$
\mathbf{E}^{\left(k+1\right)}=\mathbf{\widetilde{A}}\mathbf{E}^{\left(k\right)}=\mathbf{\widetilde{A}}^{\left(k+1\right)} \mathbf{E}^{\left(0\right)},
$}
\label{eq:no2}
\end{equation}
where $\mathbf{E}^{(k)} \in \mathbb{R}^{(N_u+N_i) \times T}$ is the \(k\)-th layer node embedding matrix, $T$ is the embedding size, $N_u$ and $N_i$ denote the number of users and items, $\mathbf{\widetilde{A}} \in \mathbb{R}^{(N_u+N_i)\times (N_u+N_i)} $ is the normalization of the adjacent matrix $\mathbf{A}$ of  the  user-item graph with self-loop, i.e., 
\begin{equation}
\begin{split}
\mathbf{\widetilde{A}}=\mathbf{\hat{D}}^{-\frac{1}{2}} \mathbf{A} \mathbf{\hat{D}}^{-\frac{1}{2}},~~
\scalebox{1}{$\mathbf{A}=\left(\begin{array}{cc}
\mathbf{I}_{N_u} & \mathbf{R} \\
\mathbf{R}^{T} & \mathbf{I}_{N_i}
\end{array}\right)$},
\end{split}
\label{eq:no1}
\end{equation}
where $\mathbf{R} \in \mathbb{R}^{N_u \times N_i}$ is the user-item interaction matrix, 
each entry $\mathbf{R}_{ui}$ is 1 if user $u$ has interacted with item $i$ otherwise 0. $\mathbf{I}_{N_u}$ and $\mathbf{I}_{N_i}$ are identity matrics with sizes of $N_u$ and $N_i$, respectively.
$\mathbf{\hat{D}}$ is a \((N_u+N_i) \times (N_u+N_i)\) diagonal matrix, in which each entry $\mathbf{\hat{D}}_{ij}$ denotes the number of nonzero entries in the \(i\)-th row vector of the adjacency matrix $\mathbf{A}$.

Note that, the initial definition of the LightGCN layer has removed the self-loop connections on nodes.   However, as the original paper claims, the layer combination operation plays a similar role by summing the weighted node embeddings at each layer as the final output representation. Along this line, we can rewrite the above message passing process for user embedding $\mathbf{e}_u$ and item  embedding $\mathbf{e}_i$  as follows,
\begin{equation}
\begin{array}{l}
\mathbf{e}_{u}^{(k+1)}=\frac{\mathbf{e}_{u}^{(k)}}{d_{u}+1}+\sum_{i \in V_{u}} \frac{\mathbf{e}_{i}^{(k)}}{\sqrt{(d_{u}+1)(d_{i}+1)}}, \\
\mathbf{e}_{i}^{(k+1)}=\frac{\mathbf{e}_{i}^{(k)}}{d_{i}+1}+\sum_{i \in V_{i}} \frac{\mathbf{e}_{u}^{(k)}}{\sqrt{(d_{i}+1)(d_{u}+1)}},
\end{array}
\label{eq:agg}
\end{equation}
where $V_u$ and $V_i$ denote the neighbor node set of user $u$ and item $i$ with size of $d_u$ and $d_i$, respectively.  The readout node embeddings $\mathbf{e}_u$ and  $\mathbf{e}_i$ are the combination of embeddings outputted by each message passing layer.
Then, the model prediction is defined as the inner product of user and item final embedding representations:
\begin{equation}
\hat{y}_{u,i}={\mathbf{e}_u}^{\intercal}{\mathbf{e}_i},
\label{eq:no6}
\end{equation}
which is used as the ranking score for recommendation.
Meanwhile, the Bayesian Personalized Ranking (BPR) loss is employed, which is a pairwise loss that encourages the possibility of an observed interaction to be higher than its unobserved counterparts, i.e.,
\begin{equation}
\mathcal{L}_{\mathrm{BPR}} = -\sum_{(u, i, j) \in \mathcal{D}} \log \sigma(\hat{y}_{ui} - \hat{y}_{uj}),
\label{eq:no7}
\end{equation}
where $\sigma(\cdot)$ is the sigmoid function,$\mathcal{D}=\{(u, i, j) \, | \, R_{u,i} = 1, R_{u,j} = 0\}$ denotes to the pairwise training dataset, and \(j\) denotes the sampled item that user \(u\) has not interacted with.


\vspace{-2mm}
\subsection{Over-Smoothing Point}
\label{sec:oversmoothing}
However, the GCN-based recommendation usually suffers performance degradation as the number of message passing layers $k$ increases due to the well-known over-smoothing problem~\cite{he2020lightgcn, wang2019neural}.
Specifically, the node embedding $\mathbf{e}^{k}$ (the row of $\mathbf{E}^{(k)}$) will all converge to the corresponding final point when \(k\) tends to infinity, called the \textbf{over-smoothing point}.
Theoretically according to Theorem 1 in \cite{chen2020simple}, we can derive that:
\begin{equation}
\mathbf{\widetilde{A}}_{u, i}^{\infty} = \mathbf{\lim} _{k \rightarrow \infty}\mathbf{\widetilde{A}}_{i, j}^{k}=\frac{\sqrt{\left(d_{u}+1\right)\left(d_{i}+1\right)}}{2N_{edge}+N_i+N_u},
\label{eq:no3}
\end{equation}
where $d_u$,$d_i$ represent the degrees of user node \(u\) and item node \(i\), respectively. $N_{edge}$ is the total number of edges in the user-item graph.
We can then obtain Over-smoothing Steady State Matrix $\mathbf{M}$ via:
\begin{equation}
\begin{split}
\mathbf{M}=\mathbf{\lim} _{k \rightarrow \infty}\mathbf{E}^{\left(k\right)}
=\mathbf{\widetilde{A}}^{\infty} \mathbf{E}^{\left(0\right)},
\end{split}
\label{eq:no4}
\vspace{-4mm}
\end{equation}
where we denote the \textbf{over-smoothing point} as $\mathbf{m}$, which is the row of $\mathbf{M}$. According to \cite{chen2020simple}, the rows of $\mathbf{M}$ are linearly correlated with each other. In other words, the high-order power of normalized adjacent matrix $\mathbf{\widetilde{A}}$ tends to undergo a rank collapse as $k$ increases. 

Actually, in the context of recommendation, the over-smoothing problem can be more serious due to the long tail and the high sparsity. On the one hand, in the item-user graph, most interactions occur on the popular users or items, which have a higher degree than most common nodes. As a result, with Equation~\ref{eq:no4}, those nodes' embeddings would converge at a higher rate and have a stronger impact on pushing the convergence of other nodes' embeddings. On the other hand, the high sparsity makes nodes interact more with local neighborhoods, due to limited direct user-item interactions. As a result, the node embeddings prefer to converge more rapidly.


\begin{figure}[t]
  \centering
    \includegraphics[width=0.50 \linewidth]{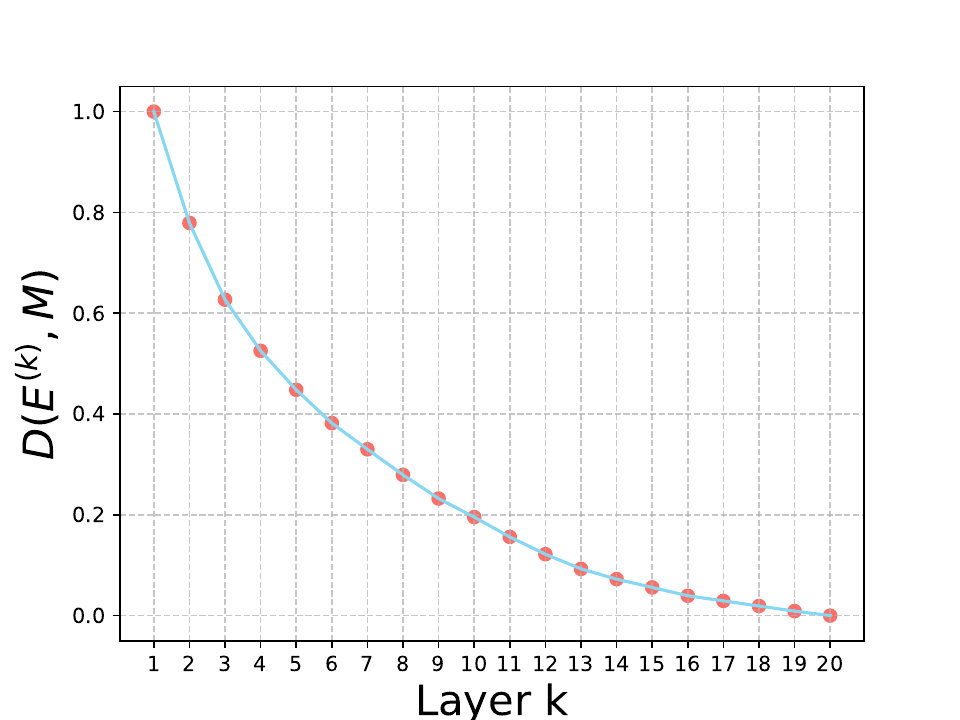}
  \hfill
    \includegraphics[width=0.45 \linewidth]{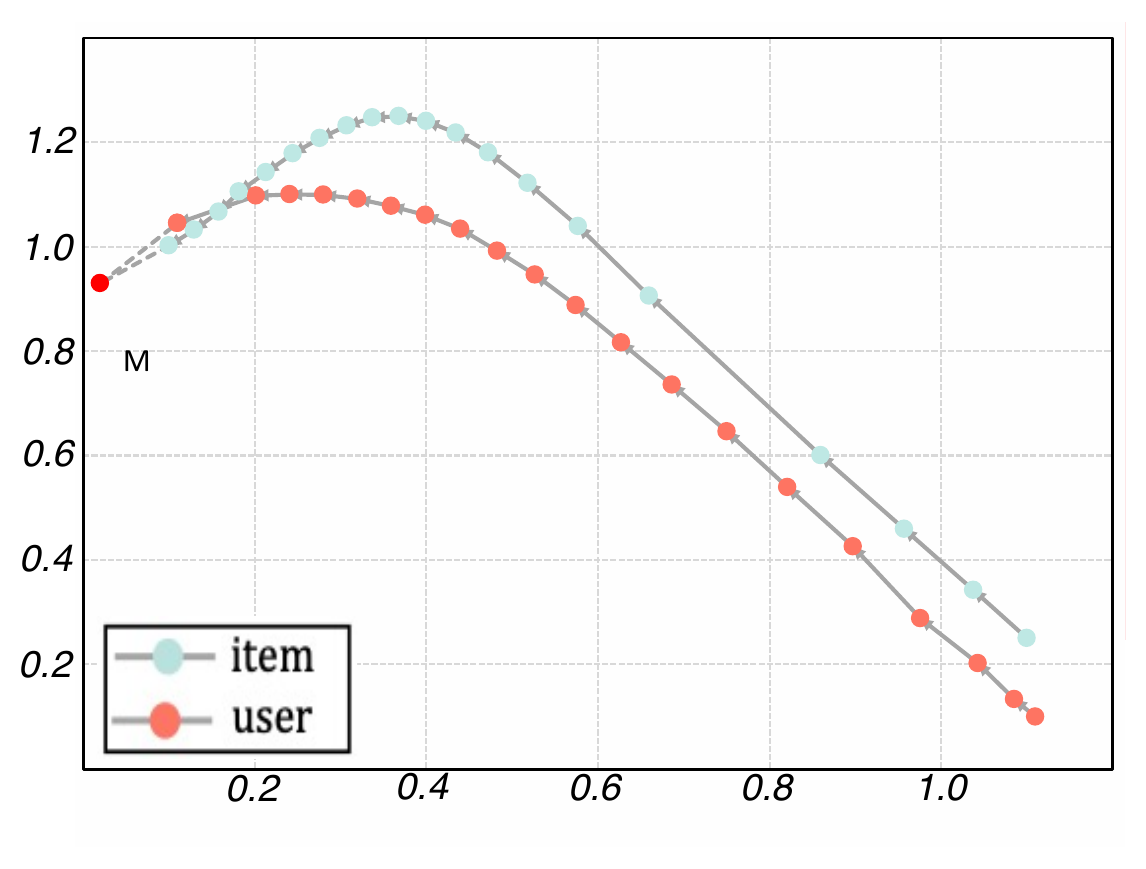}
 \vspace{-3mm}
  \caption{Empirical analysis on the evolution of the node embeddings as the layer number of message-passing increases. The left part shows the $\mathbf{D}(\mathbf{E}^{(k)},\mathbf{M})$ changes across the layer number 
\(k\) in the range [1, 20]. The right part shows the t-SNE visualization of the over-smoothing point $\mathbf{m}$ and the trajectory of embeddings of an item and a user.}
  \vspace{-4mm}
  \label{over_vis}
\end{figure}

To provide visual insights of the evolution of the node embeddings as the layer number of message-passing increases, we conduct an empirical analysis on MovieLens1M dataset in Figure \ref{over_vis} to illustrate the evolution of user or item embeddings during the message passing process. Specifically, we calculate the distance between all node embeddings and over-smoothing steady state matrix as follows:
\begin{equation}
\mathbf{D}\left(\mathbf{E}^{(k)},\mathbf{M}\right)=\frac{1}{n} \sum_{i\in[n]}\left\|\mathbf{e}_{i}^{(k)}-\mathbf{m}_{i}\right\|_{2},
\end{equation}
where $\mathbf{D}\left(\mathbf{E}^{(k)},\mathbf{M}\right)$ is the mean Euclidean distance between the \(k\)-th layer node embeddings $\mathbf{E}^{(k)}$ and over-smoothing steady state matrix $\mathbf{M}$. The $\mathbf{e}_{i}^{(k)}$ and $\mathbf{m}_{i}$ are the \(i\)-th row of $\mathbf{E}^{(k)}$ and $\mathbf{M}$ respectively. The left side of Figure \ref{over_vis} shows a rapidly decreasing trend as the layer \(k\) increases, indicating that node embeddings tend to converge towards the over-smoothing point as the layer number of message-passing increases.
Moreover, the t-SNE visualization elucidates the evolution trajectory of the embedding vector, using a user $u$ and an item $i$ as exemplars. It becomes evident that their embeddings $\mathbf{e}_i^{(k)}$ and $\mathbf{e}_u^{(k)}$ progressively converge towards the over-smoothing point \textbf{m} through the layer \(k\), even in the absence of any interaction between them. This convergence phenomenon poses a risk of recommendation failure.



\vspace{-2mm}
\section{Our Framework}
In this section, we present the technical details of our proposed \textbf{D}esmoothing Framework for \textbf{G}CN-based \textbf{R}ecommendation Systems (\textbf{DGR}). 
As shown in Figure \ref{figure1}, our DGR model is a model-free approach and can be exploited for enhancing various GCN-based recommendation systems by plugging into two components: \textbf{G}lobal Desmoothing \textbf{M}essage \textbf{P}assing (\textbf{GMP}) and \textbf{L}ocal Node \textbf{E}mbedding \textbf{C}orrection (\textbf{LEC}), which prevents global user and item embeddings from over-smoothing point but maintain similarities among their neighbor nodes. 
In the following, we will introduce those two components respectively, with a necessary analysis of the complexity.

\begin{figure}[t]
  \centering
  \includegraphics[width=0.66\linewidth]{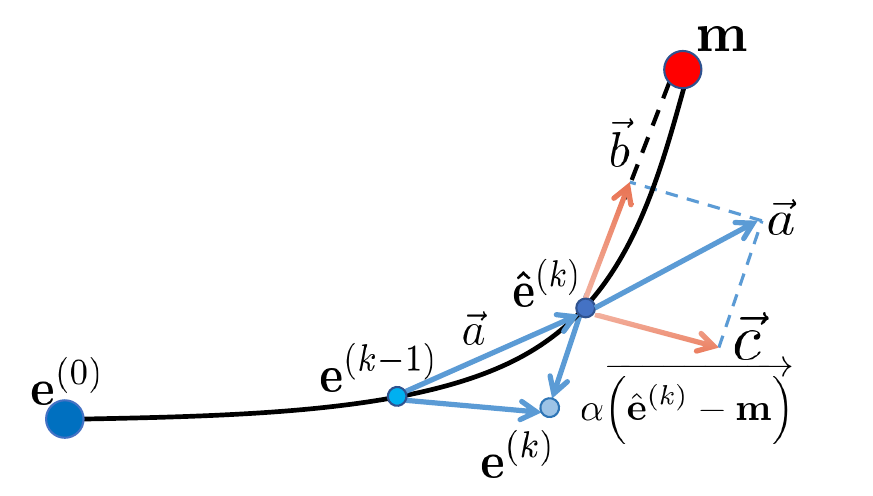}
  \vspace{-3mm}
  \caption{Illustration of the embedding vector updating in Global Desmoothing Message Passing. The solid black line from $\mathbf{e}^{(0)}$ to $\mathbf{m}$ represents the original trajectory of the GCN.}
  \label{figure2}
  \vspace{-4mm}
\end{figure}

\vspace{-1mm}
\subsection{Global Desmoothing Message Passing}\label{AA}
As discussed in Section~\ref{sec:oversmoothing}, due to the mathematical nature of global topological structure, the embedding $\mathbf{e}^{(k)} \in \mathbf{E}^{(k)}$ of each user or item will converge to the corresponding over-smoothing point $\mathbf{m} \in \mathbf{M}$ as stacking GCN layers. As a result, each user or item embedding is linearly correlated with each other and cannot be distinguished. 
To prevent the over-smoothing problem, an effective way is to slow down the convergence rate during each message passing step. Along this line, we propose a simple but effective plug-in module after each message passing layer to perturb the node embeddings away from the final over-smoothing point. 
Figure \ref{figure2} provides an illustrative understanding of our proposed approach.

\begin{figure*}[h]
  \centering  \includegraphics[width=1.0\textwidth]{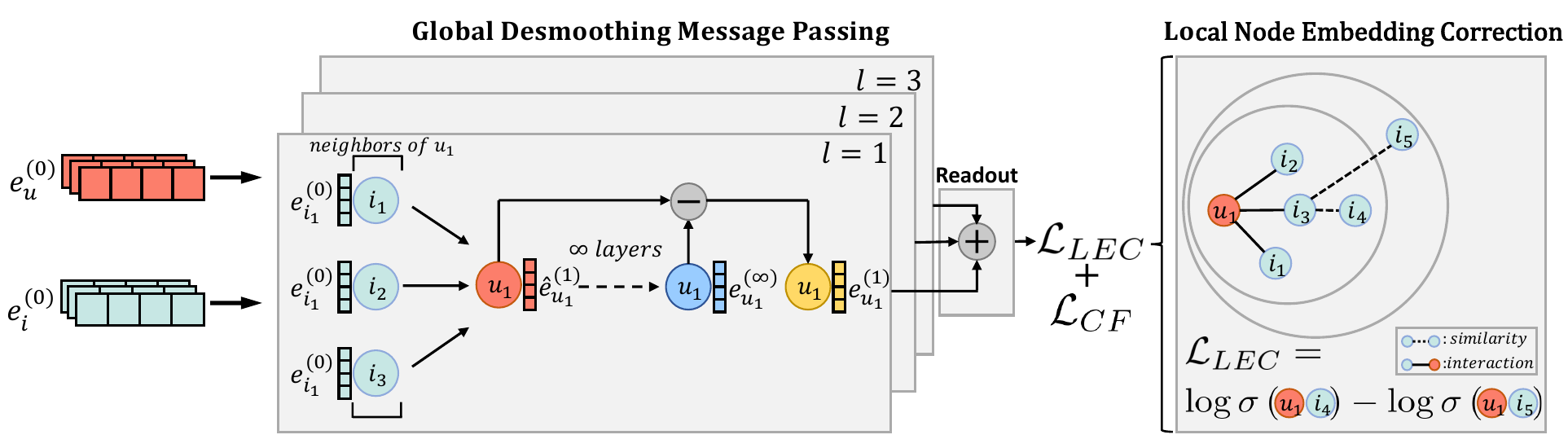}
  \vspace{-6mm}
  \caption{An illustration of DGR architecture. In the Global Desmoothing Message Passing component, user and item embeddings are put away from the over-smoothing point to preserve node embedding distinctiveness. In the Local Node Embedding Correction component, similar neighbor nodes like node $i_4$ are aggregated together while the marginal nodes like node $i_5$ is segregated in the local graph to utilize the collaborative signals. In the readout phase, $\mathbf{e}^{(0)}$ are also utilized, although they are not shown in the diagram for clarity.}
  \vspace{-3mm}
  \label{figure1}
\end{figure*}

Specifically, the black curve represents the updating trajectory of an embedding vector for a user or an item, approximating the over-smoothing point (the red point). After the \(k\)-th original message passing steps, the node embedding evolves from $e^{(k-1)}$ to $\hat{e}^{(k)}$ as Equation~\ref{eq:agg} . We denote the updating vector as $\vec{a} = \hat{e}^{(k)} - e^{(k-1)}$. It can be decomposed into two orthogonal components $\vec{b}$ and $\vec{c}$, where $\vec{b}$ is parallel to the direction of  $e^{k-1} - m$ and $\vec{c}$ is perpendicular to it. Intuitively, $\vec{b}$ pushes the embedding vector to the corresponding over-smoothing point. Therefore, we punish the vector updating in this direction with a parameter $\alpha$:
\begin{equation}
\begin{split}
\mathbf{e}^{(k)}=\hat{\mathbf{e}}^{\left(k \right)}-\alpha(\mathbf{m}-\hat{\mathbf{e}}^{\left(k\right)})=(1+\alpha)\hat{\mathbf{e}}^{\left(k \right)}-\alpha \mathbf{m},\\
\hat{\mathbf{e}}^{(k)}=AGG\left(\mathbf{e}_{j}^{(k-1)},\left\{ j \in \mathcal{N}\right\}\right),
\end{split}
\label{eq:gdmp1}
\end{equation}
where $\mathcal{N} $ denotes the set of neighbor items or users, and $AGG$ denotes the aggregation operation defined in Equation~\ref{eq:agg}. Note that $\vec{b}$ and $\mathbf{m}- \hat{\mathbf{e}}^{(k)}$ have the same direction (without sign) but different scales. Here, we use the latter to perturb the node embedding, abstaining from the use of matrix orthogonal decomposition in the computation, which enhances both robustness and usability.

Our approach to desmooth during message passing is characterized by its simplicity and intuitive nature. In practice, it has demonstrated remarkable effectiveness. 
Actually, we can also validate the effectiveness with the following proposition.
\begin{proposition} 
With the plug-in module defined in Equation~\ref{eq:gdmp1}, the distance between any node embedding $\mathbf{e}$ and the corresponding over-smoothing point $\mathbf{m}$ would increase, i.e,
\begin{equation}
    \begin{split}
        ||\mathbf{e}^{(k)} - \mathbf{m}||_p >|| \hat{\mathbf{e}}^{(k)}- \mathbf{m}||_p,
    \end{split}\label{eq:prop}
\end{equation}
where $|| \cdot||_p$ is the $p$-norm distance between vectors. 
\end{proposition}

The proof is obvious by bringing Equation~\ref{eq:gdmp1} into Equation~\ref{eq:prop}. 
The above inequality suggests that our approach prevents the collapse of user or item embeddings into the over-smoothing point by increasing the distance between them.

\vspace{-1mm}
\subsection{ Local Node Embedding Correction}\label{AA}
 In this subsection, we further focus on how to deal with the over-smoothing problem from the local graph perspective. Differently from the global view, the similarities between some local neighbor nodes are beneficial for collaborative filtering. 
It is challenging to maintain collaborative relations from local user-
item interactions while maintaining distinctiveness among all node embeddings. 

Here, we propose a general additional loss for GCN-based recommendations to handle the local user and item embedding learning process. Intuitively, the main idea is to preserve the similarity of the readout embeddings of highly neighbor and correlated item and user nodes. 
Meanwhile, The borders between similar nodes and unrelated nodes are referred to as the boundary nodes. Previous studies \cite{abbasnejad2020counterfactual,goyal2019counterfactual,wang2021counterfactual} have shown that samples in boundary nodes are often discriminative in revealing underlying data patterns, offering the potential for improved model performance. Therefore, we also hope to increase the divergence among those of boundary nodes.
As shown in Figure \ref{figure1}, given a positive pair of the user $u$ (like $u_1$) and the clicked item $i$ $(i_3)$, we first sort other neighbor item nodes $i'$   $(i_4,i_5)$ based on the distance metric between two items $i$ and $i'$. Items with lower distances are considered to have higher correlation (similar nodes $s_{im}$) and nodes with higher distances are considered marginal nodes $m_{ar}$. The numbers of similar nodes and marginal nodes are set as $K_1$ and $K_2$. Then, the additional loss term is defined as follows:
\begin{equation}
    \resizebox{.99\linewidth}{!}{$
            \displaystyle
            \begin{split}
            & \mathcal{L}_{LEC}= \\
            &\sum\limits_{(u, i)\in N^{+}} \sum\limits_{s_{im} \in S(i)}\sum\limits_{m_{ar} \in M(i)}
            -\left(\omega_{u, s_{im}} \log \sigma\left(e_{u}^{\top} e_{s_{im}}\right)-
            \omega_{u, m_{ar}} \log \sigma\left(e_{u}^{\top} e_{m_{ar}}\right)\right),\\
            &\omega_{u, s_{im}}=\frac{1}{\sqrt{d_{u}+1} \sqrt{d_{s_{im}}+1}}, ~~~
            \omega_{u, m_{ar}}=\frac{1}{\sqrt{d_{u}+1} \sqrt{d_{m_{ar}}+1}}, \\
            \end{split} $}
\label{eq:no17}
\end{equation}%
where $N^{+}$ denotes the positive pairwise training dataset, $S(i)$ is the set of similar nodes about node $i$, $M(i)$ is the set of marginal nodes about node $i$, and $\omega_{u, s_{im}}$, $\omega_{u, m_{ar}}$ are the normalization coefficients. In particular, when sorting the items, we filter out the items far enough away from the node $i$  whose distance is above a threshold $\theta$ to only preserve boundary nodes inspired by \cite{wang2021counterfactual}.
Here, we apply the co-occurrence similarity as the distance metric between two items, which is proportional to the number of users linked to both two items. Note that, the co-occurrence similarity is consistent with the idea of collaborative filtering \cite{wang2020m2grl,chen2021conet}. Hence the loss can be effectively adopted and added to common GCN-based collaborative filtering models.
The overall algorithm process of our LEC loss can be found in Appendix. 

With our proposed LEC,
we finally derive the following training objective for GCN-based recommendation models:
\begin{equation}
    \mathcal{L}=\mathcal{L}_{CF}+\lambda \mathcal{L}_{LEC},
    \label{eq:lec}
\end{equation}
where $\mathcal{L}_{CF}$ represents the original loss function of the GCN-based recommendation model like BPR loss in Equation \ref{eq:no7}, and hyper-parameter $\lambda$ is to balance the LEC loss.



\vspace{-2mm}
\subsection{Comparison with Existing Methods}
Recently, several works in the field of GCNs have been proposed to alleviate the over-smoothing problem. Here, we discuss the relations between our approach with them to provide a more comprehensive understanding.

\noindent\textbf{Residual Connection.} Motivated by the success of ResNet \cite{he2016deep}, the residual connection has been the common sense to make GNNs go deeper and perform better, like LT-OCF \cite{choi2021lt}, LightGCN \cite{he2020lightgcn}, SGC \cite{sgl}, DeepGCN \cite{li2019deepgcns}, and GCNII \cite{chen2020simple}.
Taking GCNII as an example, it incorporates initial node embeddings into the output of each layer, which can be succinctly summarized as:
\begin{equation}
\begin{split}
\mathbf{e}^{(k)}
          =&\hat{\mathbf{e}}^{\left(k \right)}+\alpha(\mathbf{e}^{(0)}-\hat{\mathbf{e}}^{\left(k \right)}),
\end{split}
\label{eq:no13}
\end{equation}
where $\hat{\mathbf{e}}^{\left(k \right)}$ and $\mathbf{e}^{(0)}$ are defined similarly to that in Equation~\ref{eq:gdmp1}. 
Intuitively, like our GMP approach, GCNII also perturbs the original node embedding $\hat{\mathbf{e}}^{\left(k \right)}$  with another vector $(\mathbf{e}^{(0)}-\hat{\mathbf{e}}^{\left(k \right)})$ of the direction moving away from over-smoothing point. The illustration of detailed embedding vector updating can be found at Appendix, which is considered to be a hidden reason that GCNII can also prevent embeddings from excessively converging to the over-smoothing point.


\noindent\textbf{Normalization-based Methods.} A proven way to reduce the over-smoothing effect is to normalize node embeddings to preserve the distinctiveness of node embeddings during training. EGNN\cite{zhou2021dirichlet} and Parinorm\cite{zhao2019pairnorm} measure over-smoothing using Dirichlet energy and Total Pairwise Squared Distance(TPSD) respectively and optimize a GCN within a constrained range of Dirichlet energy and TPSD to preserve the distinctiveness of node embeddings individually. 
Additionally, Differentiable Group Normalization (DGN)~\cite{zhou2020towards} and NodeNorm~\cite{zhou2021understanding} use node-wise normalization. Although they maintain the embedding distances between nodes to avoid excessive similarity among global node embeddings in the optimization process, these methods of optimizing node embeddings lack any directionality and can even introduce noise that affects the performance of shallow GCN models.
DGR maintains node embedding distinctiveness while providing a clear direction, which is to move away from over-smoothing point. Moreover, its application to shallow GCN models also leads to substantial improvements while normalization-based methods lead to a performance decrease.


\noindent\textbf{Topology-based Methods.} Some other approaches propose to modify the topological structure in each message passing layer to alleviate over-smoothing by randomly dropping edges or nodes in graphs, such as DropEdge\cite{rong2019dropedge} and AdaEdge\cite{chen2020measuring}. 
However, those approaches also damage the local topological structure around each node to some extent, which is critical in recommendations for modeling collaborative signals.
In contrast, our approach not only penalizes the tendency of node embeddings approximating overly to be similar, but also maintains the topological structure and similarities among neighbor nodes.
Consequently, our method yields a more stable performance enhancement for recommendation. Another topology-based model IMP-GCN \cite{liu2021interest} groups users with similar interests based on topological relationships and conducts separate message passing for each group to alleviate over-smoothing problem, which has the similar main idea with our proposed LEC, encouraging similar nodes to have comparable embeddings on the local graph.

\vspace{-2mm}
\subsection{Complexity Analysis}




Actually, our approach is easy-to-use and model-free, which can be exploited into various GCN-based recommendation systems with a limited complexity cost. 
 Assuming the original LightGCN backbone, the computational complexity can be represented as: $\mathcal{O}((N_u+N_i) \bar{d}  T  K )$, where $T$ is the dimensionality of node embeddings, $K$ is the number of graph convolutional layers, and $\bar{d}$ is the average number of neighboring nodes. After applying our DGR, the computational complexity becomes: $\mathcal{O}((N_u+N_i) (\bar{d}+1)TK)$. It can be observed that DGR only brings a limited linear increase in computational complexity. As for the space complexity, DGR only introduces additional $(N_i+N_u)T$ parameters into the original GCN backbone.
Furthermore, our experiments will demonstrate significant and robust improvements in DGR across various datasets and recommendation systems.


\renewcommand{\arraystretch}{0.8} 
\setlength{\tabcolsep}{0.8pt} 
\begin{table*}[t]
  \centering
  \vspace{-3mm}
    \begin{tabular}{ccccccccccc}
    \hline
    \toprule
    
    \multirow{2}[0]{*}{Datasets} & \multicolumn{2}{c}{Gowalla} & \multicolumn{2}{c}{MovieLens1M} & \multicolumn{2}{c}{Douban-Book} & \multicolumn{2}{c}{Yelp2018} & \multicolumn{2}{c}{Netflix} \\
          & \multicolumn{1}{c}{\textcolor[rgb]{ .02,  .388,  .757}{\small Recall@20}} & \multicolumn{1}{c}{\textcolor[rgb]{ .02,  .388,  .757}{\small NDCG@20}} & \multicolumn{1}{c}{\textcolor[rgb]{ .02,  .388,  .757}{\small Recall@20}} & \multicolumn{1}{c}{\textcolor[rgb]{ .02,  .388,  .757}{\small NDCG@20}} & \multicolumn{1}{c}{\textcolor[rgb]{ .02,  .388,  .757}{\small Recall@20}} & \multicolumn{1}{c}{\textcolor[rgb]{ .02,  .388,  .757}{\small NDCG@20}} & \multicolumn{1}{c}{\textcolor[rgb]{ .02,  .388,  .757}{\small Recall@20}} & \multicolumn{1}{c}{\textcolor[rgb]{ .02,  .388,  .757}{\small NDCG@20}} & \multicolumn{1}{c}{\textcolor[rgb]{ .02,  .388,  .757}{\small Recall@20}} & \multicolumn{1}{c}{\textcolor[rgb]{ .02,  .388,  .757}{\small NDCG@20}} \\
    \cmidrule(lr){1-11}

    BPRMF & 0.1393 & 0.1155 & 0.2486 & 0.2727 & 0.1263 & 0.1032 & 0.0496 & 0.0400 & 0.1034 & 0.0779 \\
    NGCF & 0.1602 & 0.1311 & 0.2594 & 0.2676 & 0.1509 & 0.1259 & 0.0593 & 0.0492 & 0.1056 & 0.0814 \\
    IMP-GCN & 0.1837 & 0.1510 & 0.2793 & 0.3154 & 0.1599 & 0.1349 & 0.0649 & 0.0529 & 0.1091 & 0.0795 \\
    LayerGCN & 0.1825 & 0.1502 & 0.2798 & 0.3085 & 0.1619 & 0.1363 & 0.0636 & 0.0515 & 0.1088 & 0.0787 \\
    \hline
    \toprule

    LightGCN  & 0.1832 & 0.1542 & 0.2690 & 0.3014 & 0.1497 & 0.1257 & 0.0650 & 0.0532 & 0.0868 & 0.0615 \\
    LightGCN+DGR & 0.1880 & 0.1556 & 0.2749 & 0.3065 & 0.1573 & 0.1308 & 0.0669 & 0.0545 & 0.0909 & 0.0652 \\
    \cmidrule(lr){1-11}
    \textit{Improv. } & 2.61\% & 0.92\% & 2.19\% & 1.69\% & 5.08\% & 4.06\% & 2.92\% & 2.44\% & 4.72\% & 6.02\% \\
    \hline
    \toprule

    SGL   & 0.1785 & 0.1503 & 0.2342 & 0.2194 & 0.1511 & 0.1362 & 0.0672 & 0.0553 & 0.1054 & 0.0803 \\
    SGL+DGR & 0.1823 & 0.1532 & 0.2414 & 0.2271 & 0.1625 & 0.1453 & 0.0695 & 0.0571 & 0.1105 & 0.0826 \\
    \cmidrule(lr){1-11}
    \textit{Improv. } & 2.13\% & 1.93\% & 3.07\% & 3.51\% & 7.55\% & 6.68\%  & 3.43\% & 3.26\% & 4.84\% & 2.86\% \\
    \hline
    \toprule
    SimGCL & 0.1821 & 0.1528 & 0.2560 & 0.2835 & 0.1773 & 0.1562 & 0.0717 & 0.0592 & 0.1213 & 0.0882 \\
    SimGCL+DGR & 0.1860 & 0.1573 & 0.2752 & 0.3091 & 0.1811 & 0.1601 & \underline{\textbf{0.0747}} & \underline{\textbf{0.0616}} & 0.1271 & 0.0943 \\
    \cmidrule(lr){1-11}
    \textit{Improv. } & 2.14\% & 2.95\% & 7.50\% & 9.03\% & 2.14\% & 2.50\% & 4.18\% & 4.05\% & 4.78\% & 6.92\% \\
    \hline
    \toprule
    XSimGCL & 0.1681 & 0.1388 & 0.2498 & 0.2669 & 0.1756 & 0.1590 & 0.0684 & 0.0562 & 0.1319 & 0.1014 \\
    XSimGCL+DGR & 0.1811 & 0.1533 & 0.2778 & 0.3102 & 0.1828 & 0.1633 & 0.0726 & 0.0599 & \underline{\textbf{0.1365}} & \underline{\textbf{0.1069 }}\\
    \cmidrule(lr){1-11}
    \textit{Improv. } & 7.73\% & 10.49\% & 11.21\% & 16.22\% & 4.10\% & 2.70\% & 6.14\% & 6.58\% & 3.49\% & 5.42\% \\
    \hline
    \toprule
    MixGCF & 0.1852 & 0.1579 & 0.2735 & 0.3025 & 0.1771 & 0.1606 & 0.0697 & 0.0571 & 0.1198 & 0.0858 \\
    
    MixGCF+DGR & \underline{\textbf{0.1890}} & \underline{\textbf{0.1604}} & \underline{\textbf{0.2934}} & \underline{\textbf{0.3306}} &
    \underline{\textbf{0.1922}} & \underline{\textbf{0.1730}} & 0.0719 & 0.0592 & 0.1228 & 0.0900 \\
    \cmidrule(lr){1-11}
    \textit{Improv. } & 2.05\% & 1.56\% & 7.23\% & 9.29\% & 8.53\% & 7.72\% & 3.16\% & 3.68\% & 2.50\% & 4.90\% \\
    \hline
    \toprule
    
    \end{tabular}%
    \vspace{-3mm}
    \caption{Performance comparison of baseline models before and after integrating DGR.}
  \label{tab:zhushiyan}%
  \vspace{-3mm}
\end{table*}

\vspace{-2mm}
\section{Experiments}
To verify the effectiveness of the proposed DGR, we conduct extensive experiments based on five public benchmark datasets and report detailed analysis results in the section.
\vspace{-2mm}
\subsection{Experimental Setup}
\textbf{Datasets and Evaluation.} Following many recent GCN-based recommendation systems \cite{he2020lightgcn,mao2021ultragcn,yu2023xsimgcl}, five public benchmark datasets are used in our paper including Gowalla, Yelp2018, MovieLens1M, Netflix and Douban-Book. These datasets vary in domains, scale, and density. The statistics of the datasets can be found in Appendix. 
We partitioned each dataset into training, and testing sets using
a 4:1 ratio. Recall@K and Normalized Discounted Cumulative Gain (NDCG)@K are chosen as the evaluation metrics that are popular in the evaluation of GCN-based recommendations. 

\noindent\textbf{Baseline Models.} To demonstrate the generalizability, we incorporate our DGR into five widely-use GCN-based recommendation models including LightGCN \cite{he2020lightgcn}, SGL \cite{sgl}, SimGCL \cite{yu2020graph}, XSimGCL \cite{yu2023xsimgcl} and MixGCF \cite{huang2021mixgcf}. In addition, we include two traditional recommendation models, BPRMF \cite{rendle2012bpr} and NGCF \cite{wang2019neural}. The latest state-of-the-art models, IMP-GCN and LayerGCN \cite{zhou2023layer}, are also included, which effectively mitigate the over-smoothing promblem in recommendation systems.

\noindent\textbf{Experimental Setting.} The experimental implementation details can be found in the Appendix.

\renewcommand{\arraystretch}{0.8} 
\setlength{\tabcolsep}{0.8pt} 
\begin{table*}[t]
  \centering
    \begin{tabular}{ccccccccccc}
    \hline
    \toprule
    \multirow{2}[0]{*}{Model} & \multicolumn{2}{c}{LightGCN} & \multicolumn{2}{c}{SGL} & \multicolumn{2}{c}{SimGCL} & \multicolumn{2}{c}{XSimGCL} & \multicolumn{2}{c}{MixGCF} \\
          & \multicolumn{1}{c}{\textcolor[rgb]{ .02,  .388,  .757}{\small Recall@20}} & \multicolumn{1}{c}{\textcolor[rgb]{ .02,  .388,  .757}{\small NDCG@20}} & \multicolumn{1}{c}{\textcolor[rgb]{ .02,  .388,  .757}{\small Recall@20}} & \multicolumn{1}{c}{\textcolor[rgb]{ .02,  .388,  .757}{\small NDCG@20}} & \multicolumn{1}{c}{\textcolor[rgb]{ .02,  .388,  .757}{\small Recall@20}} & \multicolumn{1}{c}{\textcolor[rgb]{ .02,  .388,  .757}{\small NDCG@20}} & \multicolumn{1}{c}{\textcolor[rgb]{ .02,  .388,  .757}{\small Recall@20}} & \multicolumn{1}{c}{\textcolor[rgb]{ .02,  .388,  .757}{\small NDCG@20}} & \multicolumn{1}{c}{\textcolor[rgb]{ .02,  .388,  .757}{\small Recall@20}} & \multicolumn{1}{c}{\textcolor[rgb]{ .02,  .388,  .757}{\small NDCG@20}} 
          \\
    \cmidrule(lr){1-11}
    Baseline Model  & 0.1497 & 0.1257 & 0.1511 & 0.1362 & 0.1773 & 0.1562 & 0.1756 & 0.1590 & 0.1771 & 0.1606 \\
    \cmidrule(lr){1-11}
    Model+GMP & 0.1533 & 0.1306 & 0.1575 & 0.1418 & 0.1786 & 0.1590 & 0.1777 & 0.1599 & 0.1807 & 0.1623 \\
    \textit{Improv. } & +2.40\% & +3.90\% & +4.24\% & 4.11\% & +0.73\% & +1.79\% & +1.20\% & +0.57\% & +2.03\% & +1.06\% \\
    \cmidrule(lr){1-11}
    Model+LEC & 0.1522 & 0.1279 & 0.1511 & 0.1363 & 0.1790 & 0.1575 & 0.1814 & 0.1631 & 0.1914 & 0.1716 \\
    \textit{Improv. } & +1.67\% & +1.75\% & +0.00\% & +0.07\% & +0.96\% & +0.83\% & +3.30\% & +2.58\% & +8.07\% & +6.85\% \\
    \cmidrule(lr){1-11}
    Model+DGR & 0.1573 & 0.1308 & 0.1625 & 0.1453 & 0.1811 & 0.1601 & 0.1828 & 0.1633 & 0.1922 & 0.1730 \\
    \textit{Improv. } & +5.08\% & +4.06\% & +7.54\% & +6.68\% & +2.14\% & +2.50\% & +4.10\% & +2.70\% & +8.53\% & +7.72\% \\
    \hline
    \toprule
    \end{tabular}%
    \vspace{-3mm}
    \caption{Performance comparison of GMP and LEC components in DGR on Douban-Book.}
  \label{tab3}%
  \vspace{-4mm}
\end{table*}

\vspace{-2mm}
\subsection{Overall Performance}
Table \ref{tab:zhushiyan} summarizes the comparison of baseline models before and after integrating the proposed DGR with the percentage of relative improvement at each metric. The main observations are as follows:

\indent \textbf{(1)} The baseline models incorporating DGR all exhibit performance improvements on all datasets, while also achieving state-of-the-art results. It demonstrates the effectiveness of the DGR, which unlocks the potential of personalized recommendation for GCN-based recommendation systems by mitigating the over-smoothing problem.
In the vast majority of cases, the five popular models with DGR significantly outperform competitive traditional and recent state-of-the-art models.
Baselines and Baselines with DGR in Table \ref{tab:zhushiyan} apply the same layers of GCNs, which demonstrates that our proposed DGR can significantly improve shallow GCN-based recommendation models.

\indent \textbf{(2)} Compared to all other models, DGR yields the substantial improvement for XSimGCL and SimGCL among all datasets. In particular, XSimGCL with DGR achieves a remarkable increase of 11.21\% in Recall@20 and 16.22\% in NDCG@20 on the MovieLens1M. We believe that XSimGCL and SimGCL employ the approach of adding noise to node embeddings instead of using complex data augmentation for graph self-supervised learning. This method may not effectively leverage topological information in the local graph, and introducing noise makes the model directionless. In contrast, DGR enables them to maintain distinctiveness among all user and item embeddings under the guidance of global topology while paying closer attention to the local topological information of local user-item interactions, leading to the enhanced performance.

\indent \textbf{(3)} Comparing the performance on different datasets, DGR exhibits the substantial  average improvement on MovieLens1M among all baseline models, which has the highest density, at 2.7\%. The DGR can improve recommendation performance remarkably by slowing down its tendency towards over-smoothing.  
Meanwhile, DGR also has stable improvements on sparse datasets such as Yelp2018 with 0.13\% density. 
This is because DGR can effectively utilize the limited neighbors of each user (item) node to capture collaborative signals by learning on the local graph.

\begin{figure}[t]
  \centering
  \includegraphics[width=0.68\linewidth]{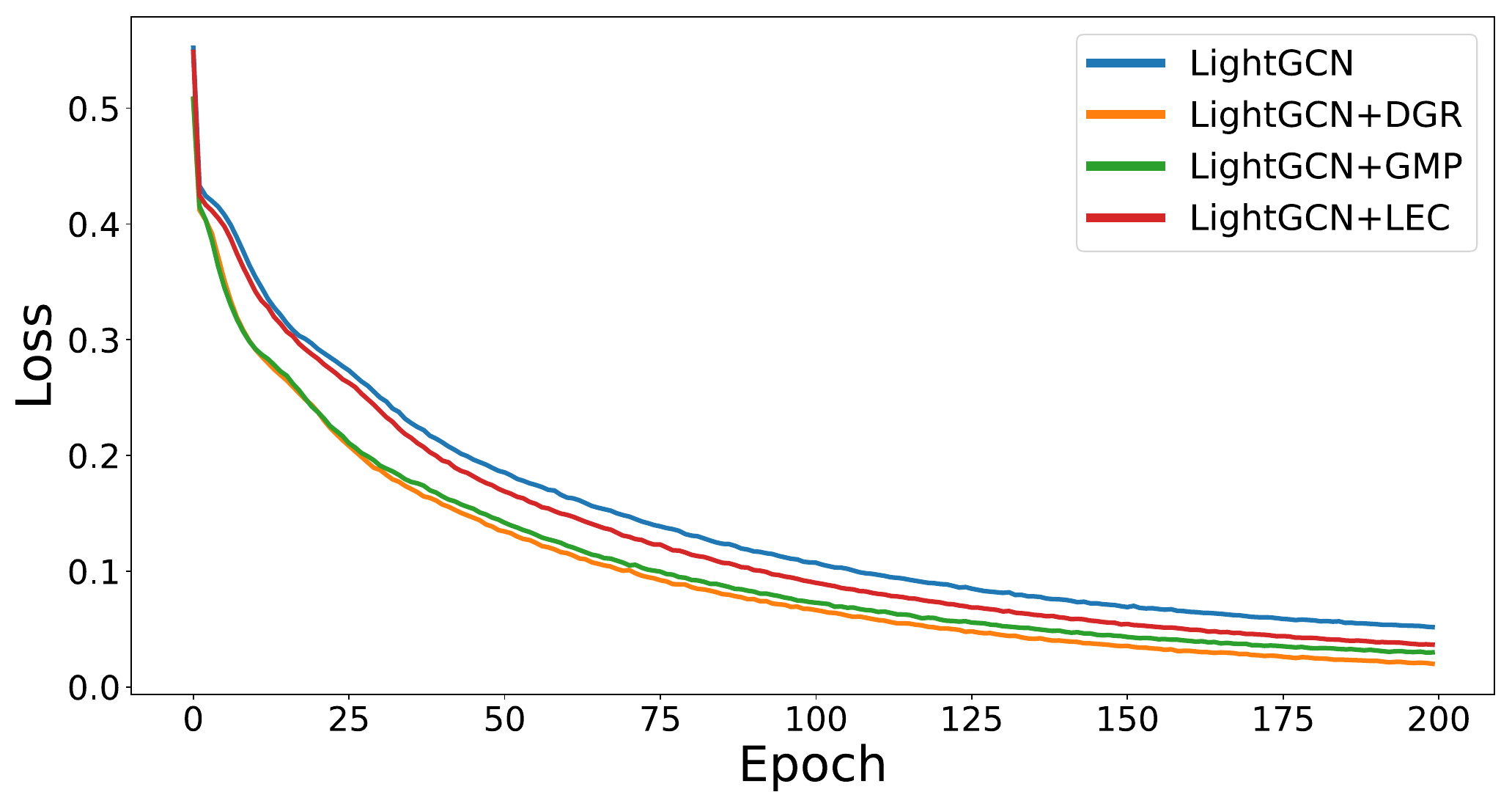}
  \vspace{-2mm}
  \caption{Training losses with epochs on Douban-Book}
  \vspace{-2mm}
  \label{lossplot}
\end{figure}

\noindent\textbf{Efficiency.} To experimentally validate the efficiency of our proposed method, we train LightGCN, LightGCN+GMP, LightGCN+LEC, and LightGCN+DGR separately on the Douban-Book dataset. We record the training loss over training epochs, as depicted in Figure \ref{lossplot}. It is evident that during the training process, our model exhibits higher training efficiency compared to the baseline model.

\vspace{-1mm}
\subsection{Further Analysis of DGR}
In this section, we provide more experimental analysis of our proposed DGR for better understanding. 

\noindent\textbf{Ablation Study of DGR.} We conduct the ablation experiments on Douban-Book by individually applying GMP, LEC, and DGR on different models. From Table \ref{tab3}, we can find that both GMP and LEC contribute to the improvements of model performance. The performance improvements provided by DGR is greater than the sum of the performance improvements from GMP and LEC, which implies that the synergistic combination of these two components is better suited for the desmoothing task in the context of recommendation.
Additionally, we observe that GMP leads to a more remarkable performance improvement than LEC in LightGCN, SGL, and SimGCL. However, this trend is reversed in XSimGCL and MixGCF. We believe the primary reason is that the first three models are already capable of effectively capturing collaborative information during the aggregation of neighboring items(users), which makes global over-smoothing their performance bottleneck. Meanwhile, the latter two models are perceived as lacking the ability to capture collaborative signals from user-item interactions. 

\begin{figure}[t]
  \centering
  \includegraphics[width=0.7\linewidth]{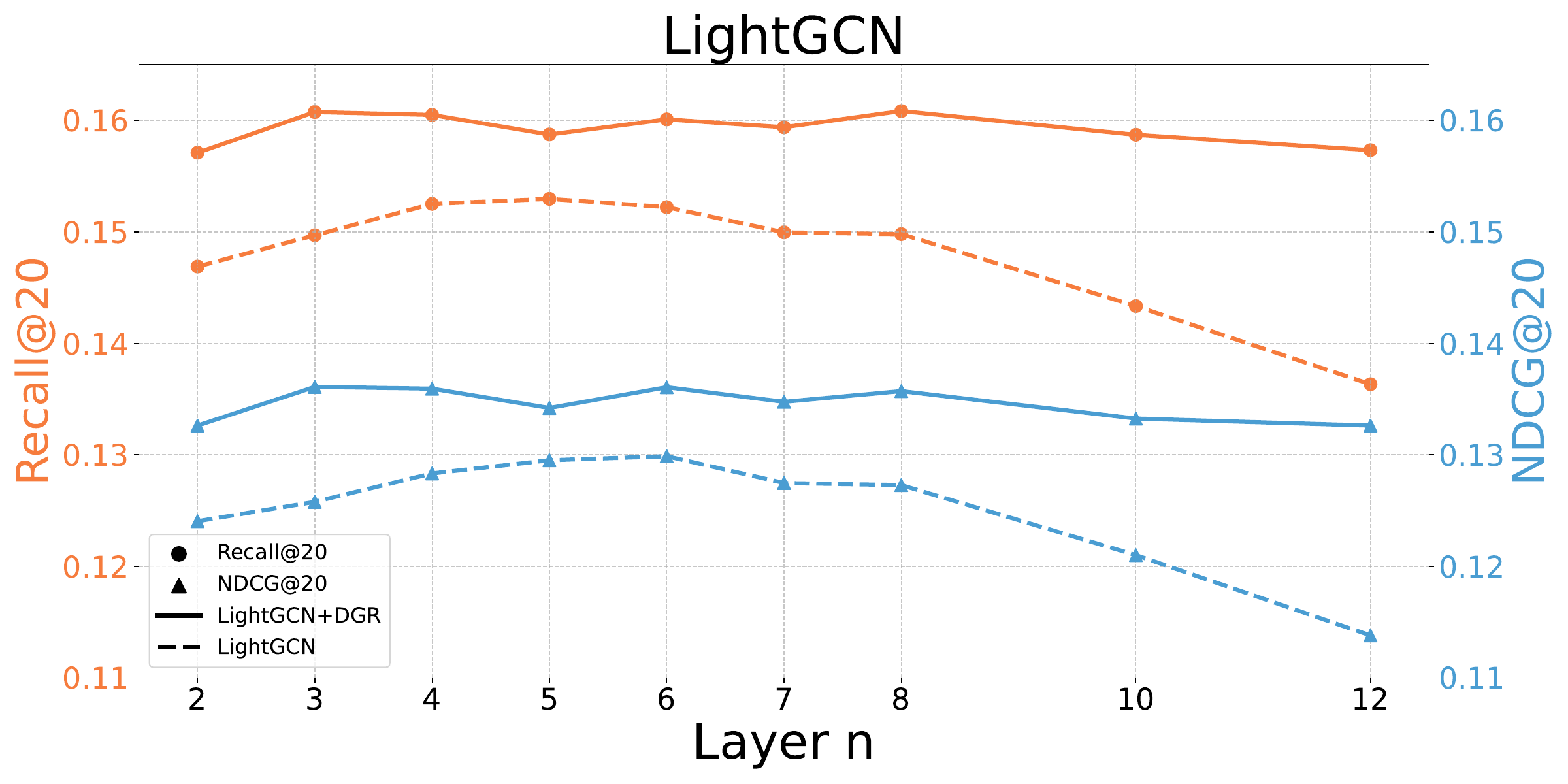}
  \vspace{-2mm}
  \caption{Performance of LightGCN with and without the DGR across a range of model depths. The orange line and blue line represent Recall@20 and NDCG@20, respectively. The solid line and dashed line represent models with and without DGR, respectively.}
  \label{deepdgr}
  \vspace{-4mm}
\end{figure}

\noindent\textbf{Case Study.} The \textbf{Row-diff} is a metric to measure the degree of over-smoothing, proposed by \cite{zhao2019pairnorm}, where a lower row-diff score indicates the larger degree of over-smoothing. Its calculation formula is as follows:
\begin{equation}
\textbf{ Row-diff }\left(\mathbf{E}^{(k)}\right)=\frac{1}{n^{2}} \sum_{i, j \in[n]}\left\|\mathbf{e}_{i}^{(k)}-\mathbf{e}_{j}^{(k)}\right\|_{2},
\label{row-diff}
\end{equation}
where $\mathbf{E}^{(k)}$ is the node embedding matrix obtained after training \(k\) layers of GCN until convergence. In order to validate the ability of DGR to mitigate over-smoothing, we separately train baseline models and baseline models with DGR until convergence to obtain a converged embedding matrix. The row-diffs of node embeddings are computed respectively and shown at Table \ref{tab:case}. It is evident that the application of DGR to the baseline models leads to a significant reduction in the degree of over-smoothing.

\renewcommand{\arraystretch}{1.0}  
\setlength{\tabcolsep}{0.8pt} 
\begin{table}[t]
\vspace{2mm}
  \centering
    \begin{tabular}{cccccc}
\hline
\bottomrule
    \small\textbf{Model} & \multicolumn{1}{c}{\small\textbf{LightGCN}} & \multicolumn{1}{c}{\small\textbf{SGL}} & \multicolumn{1}{c}{\small\textbf{SimGCL}} & \multicolumn{1}{c}{\small\textbf{XSimGCL}} & \multicolumn{1}{c}{\small\textbf{MixGCF}} \\
    \cmidrule(lr){1-6}
    \small Baseline  & 2.68 & 2.61 & 2.65 & 2.32 & 2.38\\
    \small Baseline+DGR & 2.80 & 2.85 & 3.06 & 2.75 & 2.74 \\
\hline
\bottomrule
    \end{tabular}
    \vspace{-1mm}
    \caption{\textbf{Row-diffs} of baseline models with and without DGR}
    \vspace{-2mm}
    \label{tab:case}
\end{table}

\noindent\textbf{Impact of Model Depth.} Although the DGR significantly enhances the shallow GCN-based recommendations as shown in Table \ref{tab:zhushiyan}, further exploration is needed to understand its desmoothing ability in deep GCN-based recommendations. We evaluate the five models with and without DGR across a range of model depths. We conduct these experiments on the Douban-Book dataset. The results of LightGCN are shown in Figure \ref{deepdgr} while the results of other models can be found in Appendix.
It is notable that models with DGR display a slower decline in performance as the number of layers increases, in contrast to those without DGR. This observation suggests that DGR possesses a robust desmoothing capability, which can effectively alleviate the detrimental effects of over-smoothing on recommendation performance in deep models.


\begin{figure}[t]
  \centering
    \includegraphics[width=0.32 \linewidth]{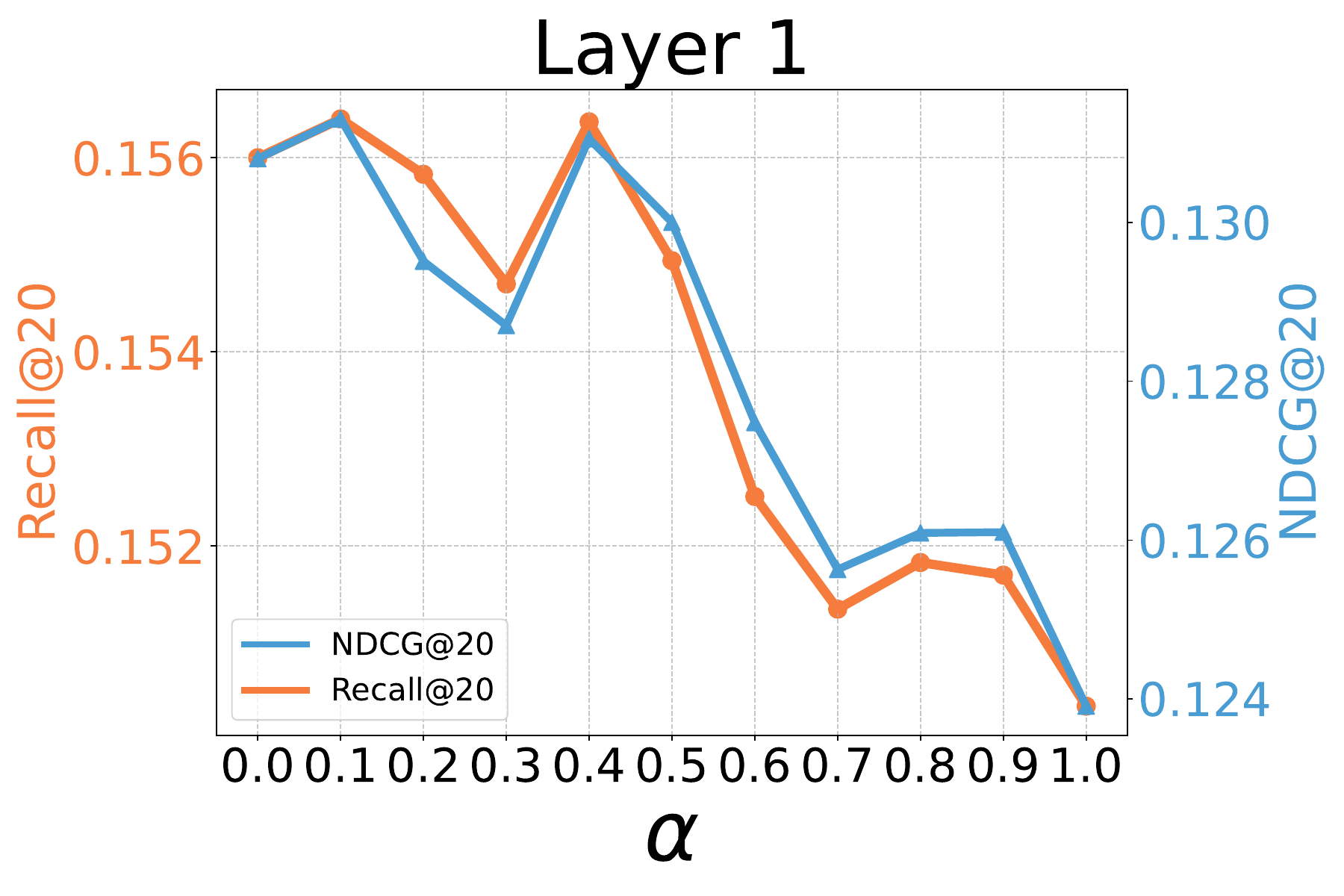}
  \hfill
    \includegraphics[width=0.32 \linewidth]{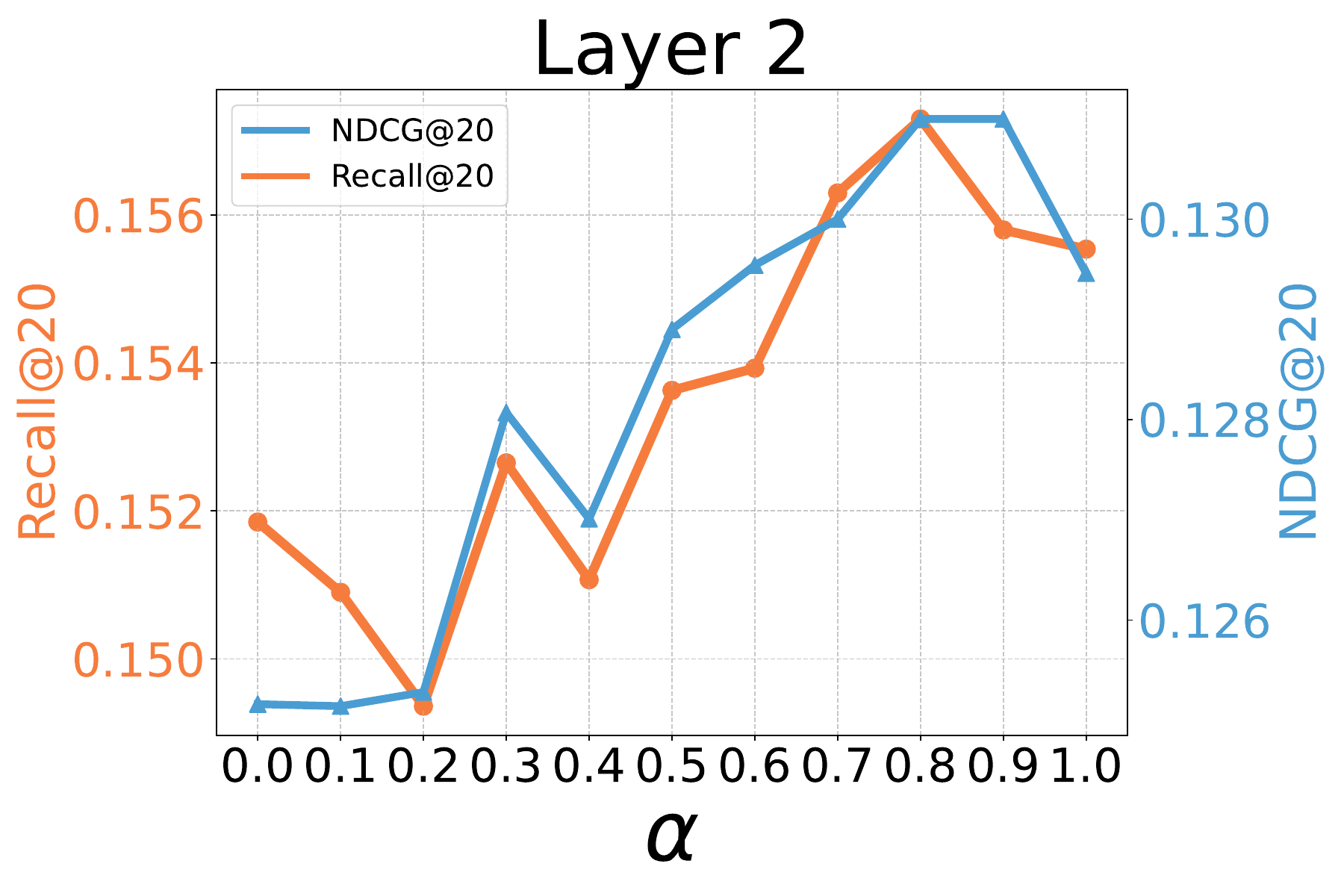}
   \hfill
      \includegraphics[width=0.32 \linewidth]{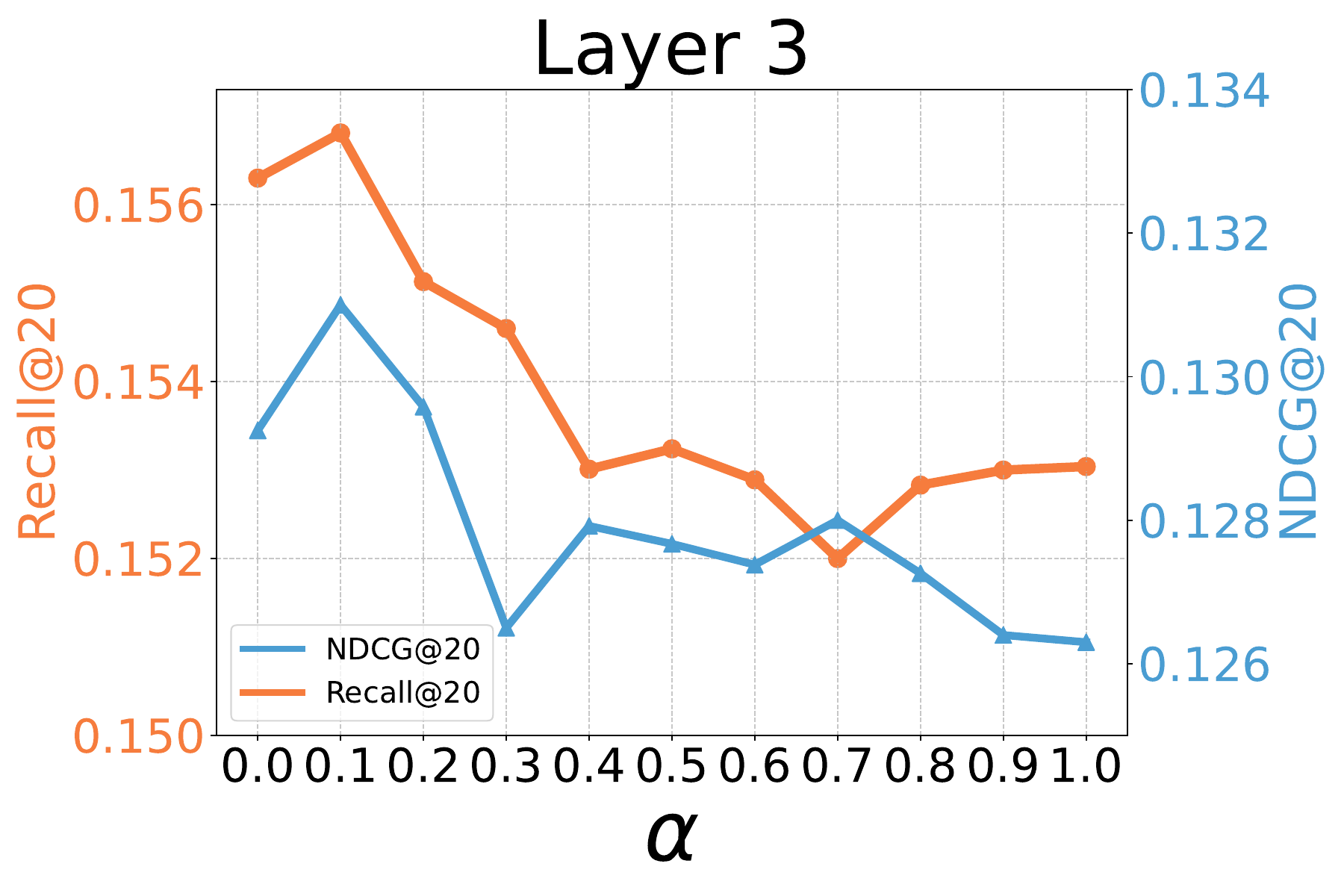}
      \vspace{-2mm}
  \caption{Parameter  Analysis of $\alpha$ based on LightGCN.}
  \vspace{-2mm}
  \label{over5}
\end{figure}

\noindent\textbf{Impact of hyper-parameters.} As mentioned in Equation \ref{eq:gdmp1}, the hyper-parameter $\alpha$ determines the extent to which all embeddings deviate from the over-smoothing point. It has a significant impact on the model performance. Here, We take LightGCN as the representation example to incorporate our DGR model with varying $\alpha$ values on Douban-Book dataset, while the results of other models can be found in Appendix.
The optimal value of $\alpha$ varies across different layers. Therefore, we explore the optimal $\alpha$ separately for each layer across a range of values, i.e., [0, 0.1, 0.2, 0.3, 0.4, 0.5, 0.6, 0.7, 0.8, 0.9, 1.0]. We set the number of message passing layers as 3.
The result for LightGCN is shown in Figure \ref{over5}. We can find the optimal $\alpha$ is 0.1 for the first layer, 0.8 for the second layer, and 0.1 for the third layer, concluded as [0.1,0.8,0.1].
Similarly, we find the optimal parameters for other models, i.e, [1.0, 0.8, 0.8] for SGL, [0, 0.6, 0.6] for SimGCL, [0.2, 0.2, 0.2] for XSimGCL and [0, 0.4, 0.2] for MixGCF. More details about other hyper-parameters in our proposed DGR are available in Appendix.

\vspace{-2mm}
\section{Conclusion}
In this work, we proposed an general and easy-to-use \textbf{D}esmoothing Framework for \textbf{G}CN-based \textbf{R}ecommendation Systems(\textbf{DGR}), to alleviate the over-smoothing problem in GCN-based recommendation systems, thereby unlocking the potential for personalized recommendation. We approached this problem from both global and local perspectives. 
Specifically, with the guidance of the global structure, we introduced \textbf{G}lobal Desmoothing \textbf{M}essage \textbf{P}assing(\textbf{GMP}), which penalizes the tendency of node embeddings approximating overly to be similar.
In addition, we introduced \textbf{L}ocal Node \textbf{E}mbedding \textbf{C}orrection (\textbf{LEC}) for the readout embeddings to preserve the local collaborative relations between users and their neighboring items in the local graph. Extensive experiments based on five public datasets and five popular GCN-based recommendation models have demonstrated the effectiveness of our proposed DGR. 

\section*{Acknowledgements}
This work was partially supported by Project funded by China Postdoctoral Science Foundation (Grant No.2023M730785).

\vspace{-2mm}
\bibliographystyle{named}
\bibliography{ijcai24}
\clearpage

\appendix

\vspace{-2mm}
\section{Related Work}
This section summarizes the additional literature related to our paper, which can be split into two aspects, Graph-based Collaborative Filtering and Over-smoothing in GCNs.

\vspace{-2mm}
\subsection{Graph-based Collaborative Filtering}
Collaborative Filtering (CF) is a widely applied recommendation system approach. CF predicts user interests by leveraging the preferences of other users with similar interests \cite{goldberg1992using}. Early memory-based CF methods predict user preferences by calculating the similarity between users or items \cite{linden2003amazon,hofmann2004latent}. Recently, model-based methods became prevalent, which represent users and items as latent vectors and compute their dot product to predict unobserved ratings \cite{breese2013empirical,su2006collaborative}. 
Subsequent work has primarily focused on two main directions: improving embeddings or optimizing interaction functions. Because interaction data can be effectively modeled as bipartite graphs, the use of neural graph networks in CF aligns well with the aforementioned requirements. Methods like NGCF \cite{ngcf}, PinSage \cite{personlized1}, LightGCN \cite{he2020lightgcn}, LR-GCCF\cite{chen2020revisiting} and LCF \cite{yu2020graph} have achieved significant success in this regard. These approaches restructure individual histories from the graph and extract valuable information from multi-hop neighbors to refine embeddings. Additionally, some studies \cite{sun2019multi} have further proposed to construct additional interaction graphs for capturing richer associations between users and items. In particular, our approach in this paper is model-free and can be adapted to most of the above GCN-based recommendations.

\vspace{-2mm}
\subsection{Over-smoothing in GCN}
Graph Convolutional Networks (GCNs) have gained extensive utilization in the modeling of real-world graphs\cite{chen2020mmea}, encompassing domains like protein networks \cite{ying2018hierarchical}, resource allocation networks \cite{tficc}, and co-author networks \cite{gcn_yuanshi}, and have yielded notable achievements. 
However, one recurrent issue frequently encountered during the training of deep GCNs is the phenomenon of over-smoothing \cite{li2019deepgcns}, where repeated graph convolutions blend neighborhood embeddings to the extent that the output of an infinitely-deep GCN converges to the same.
Recently, several approaches aimed at alleviating or resolving this issue in GCNs have achieved significant success, such as DenseGCN \cite{li2019deepgcns}, DropEdge \cite{rong2019dropedge}, PairNorm \cite{zhao2019pairnorm}, BORF \cite{nguyen2023revisiting} and DAGNN \cite{liu2020towards}.
However, those approaches may not be effective sufficiently for the GCN-based recommendation due to the bipartite and sparse nature of graphs in the recommendation, which requires the desmoothing framework to not only reduce smoothing but also capture collaborative signals more effectively. Therefore, it is still a challenging problem to design an expressive desmoothing framework for the recommendation.

\section{Illustration of GCNII the embedding vector updating}
\begin{figure}[htbp]
  \centering
  \includegraphics[width=0.3\textwidth]{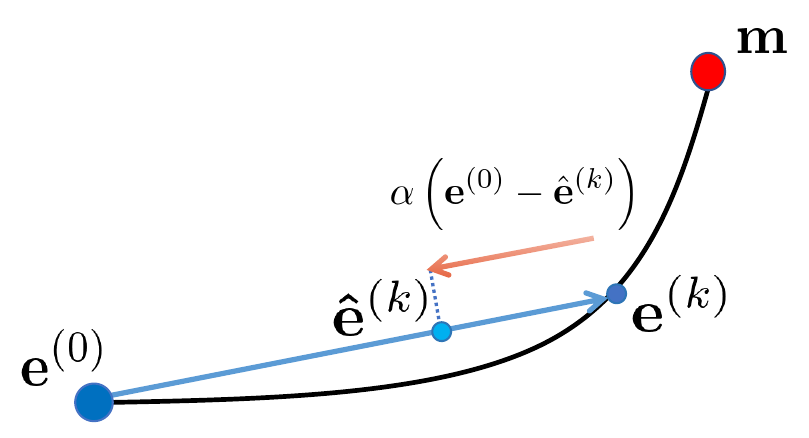}
  \caption{Illustration of the embedding vector updating in GCNII. The solid black line from $\mathbf{e}^{(0)}$ to $\mathbf{m}$ represents the original trajectory of the GCN}
  \label{figure3}
\end{figure}

From our desmoothing perspective, we have illustrated the underlying principles of the GCNII model, as shown in Figure~\ref{figure3}.

\label{app:a}
\vspace{-4mm}
\section{Experimental Implementation Detail}
\label{app:b}
In this sction, we will introduce our experimental environments and the dataset statistics.The following software and hardware environments were used for all experiments:Ubuntu 18.04.3 LTS, Python 3.7.0, Pytorch 1.13.1, Numpy 1.20.3, Scipy 1.6.2, CUDA 11.6, NVIDIA Driver 510.108.03, Intel(R) Xeon(R) Gold 6226 CPU and Tesla V100.
The embedding size $T$ for all models is 64, which has a significant impact on the performance of personalized recommendation. The learning rate is set to 1e-3. And the batch size is 2048. The dataset statistics are presented at Table~\ref{datasettable}.

\renewcommand{\arraystretch}{1.0}  
\setlength{\tabcolsep}{3.0pt} 
\begin{table}[h]
  \centering
  
    \begin{tabular}{cccccc}
\hline
\bottomrule
    \textbf{Dataset} & \multicolumn{1}{c}{\textbf{\#Users}} & \multicolumn{1}{c}{\textbf{\#Items}} & \multicolumn{1}{c}{\textbf{\#Interactions}} & \multicolumn{1}{c}{\textbf{Density}} \\
    \cmidrule(lr){1-5}
    Douban-Book  & 12,859 & 22,294 & 598,420 & 0.21\% \\
    MovieLens1M & 6,038 & 3,533 & 575,281 & 2.70\% \\
    Gowalla & 29,858 & 40,981 & 1,027,370 & 0.08\% \\
    Yelp2018 & 31,668 & 38,048 & 1,561,406 & 0.13\% \\
    Netflix & 46,420 & 12,898 & 2,475,020 & 0.41\% \\
\hline
\bottomrule
    \end{tabular}
    \vspace{-2mm}
    \caption{Basic dataset statistics.}
    \vspace{-4mm}
    \label{datasettable}
\end{table}

\vspace{-2mm}
\section{Algorithm of LEC}
\label{app:c}
The algorithm process of LEC is shown in Algorithm~\ref{alg1}.

\begin{algorithm}[htbp]
\setstretch{1.0}
      \caption{Local Node Embedding Correction}  
      \label{al:alg1}  
    \begin{algorithmic}[1]
      \STATE {\bfseries Input:} The positive pairwise training dataset $N^{+}$; The number of similar nodes $K_1$; The number of marginal nodes $K_2$; The threshold of marginal nodes $\theta$;
    \FOR{positive pair $u,i$ in $N^{+}$}
 
            \STATE Select the first-order neighbors of item $i$ as $\mathcal{N}_i$;
            \STATE Select the second-order neighbors of item $i$ as 
            $\mathcal{N}^{(2)}_i$;
            \STATE $\mathcal{Q} \leftarrow \{\}$; 
            \FOR{item $j$ in $\mathcal{N}^{(2)}_i$}
              \STATE Select the first-order neighbors of item $j$ as $\mathcal{N}_j$;
              \STATE $count_j$ $\leftarrow \left|\mathcal{N}_i \cap \mathcal{N}_j \right| $;
              \IF{$count_j \leq \theta$}
               \STATE Continue;
               \ELSE
                \STATE $\mathcal{Q} \leftarrow \{(j,count_j)\} \cup \mathcal{Q}$;
              \ENDIF
            \ENDFOR
            \STATE Sort $\mathcal{Q}$ by $count_j$;
            \STATE Select the first $K_1$ nodes from $\mathcal{Q}$ as $\mathcal{S}(i)$ by $count_j$;
            \STATE Select the last $K_2$ nodes from $\mathcal{Q}$ as $\mathcal{M}(i)$ by $count_j$;
 
   \ENDFOR
   \STATE Calculate $\mathcal{L}_{LEC}$ as Equation \ref{eq:no17};
     \STATE {\bfseries Output:} $\mathcal{L}_{LEC}$;

  \end{algorithmic}
  \label{alg1}
\end{algorithm}


\vspace{-4mm}
\section{Performance of models with and without the DGR across a range of model depths.}
\begin{figure*}[htbp]
  \centering
  \vspace{1pt}
  \includegraphics[width=0.45\textwidth]{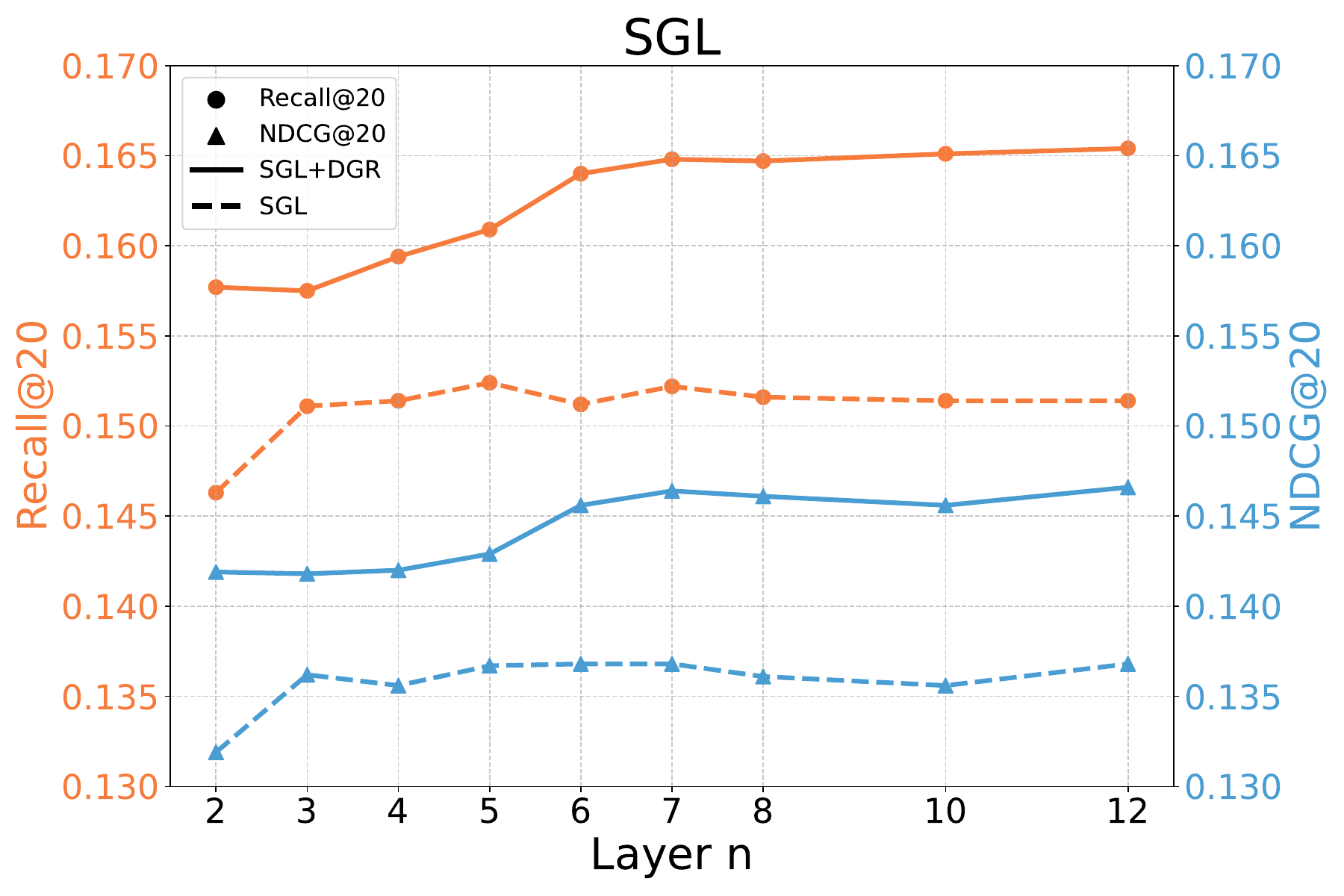}
  \includegraphics[width=0.45\textwidth]{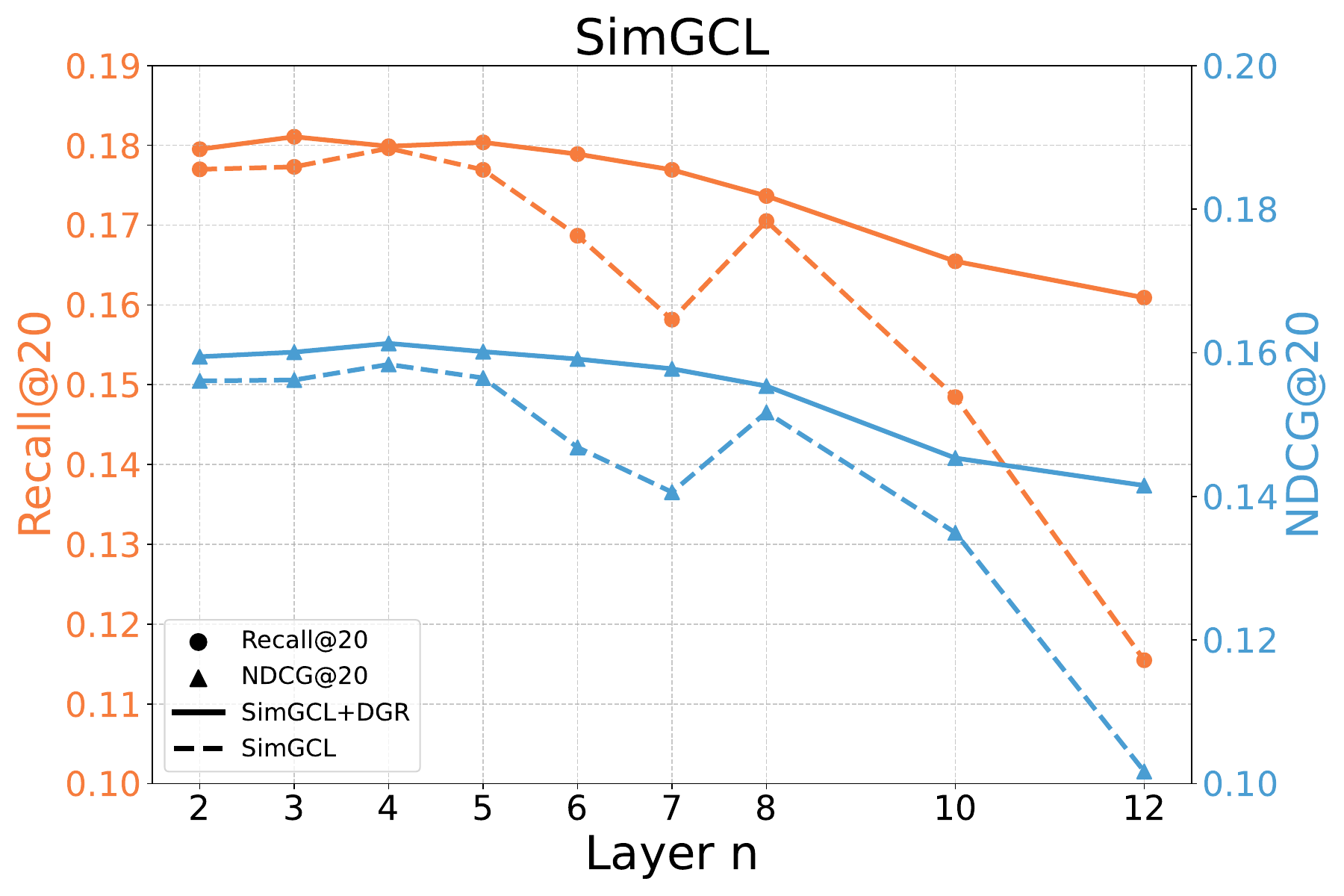}
  \includegraphics[width=0.45\textwidth]{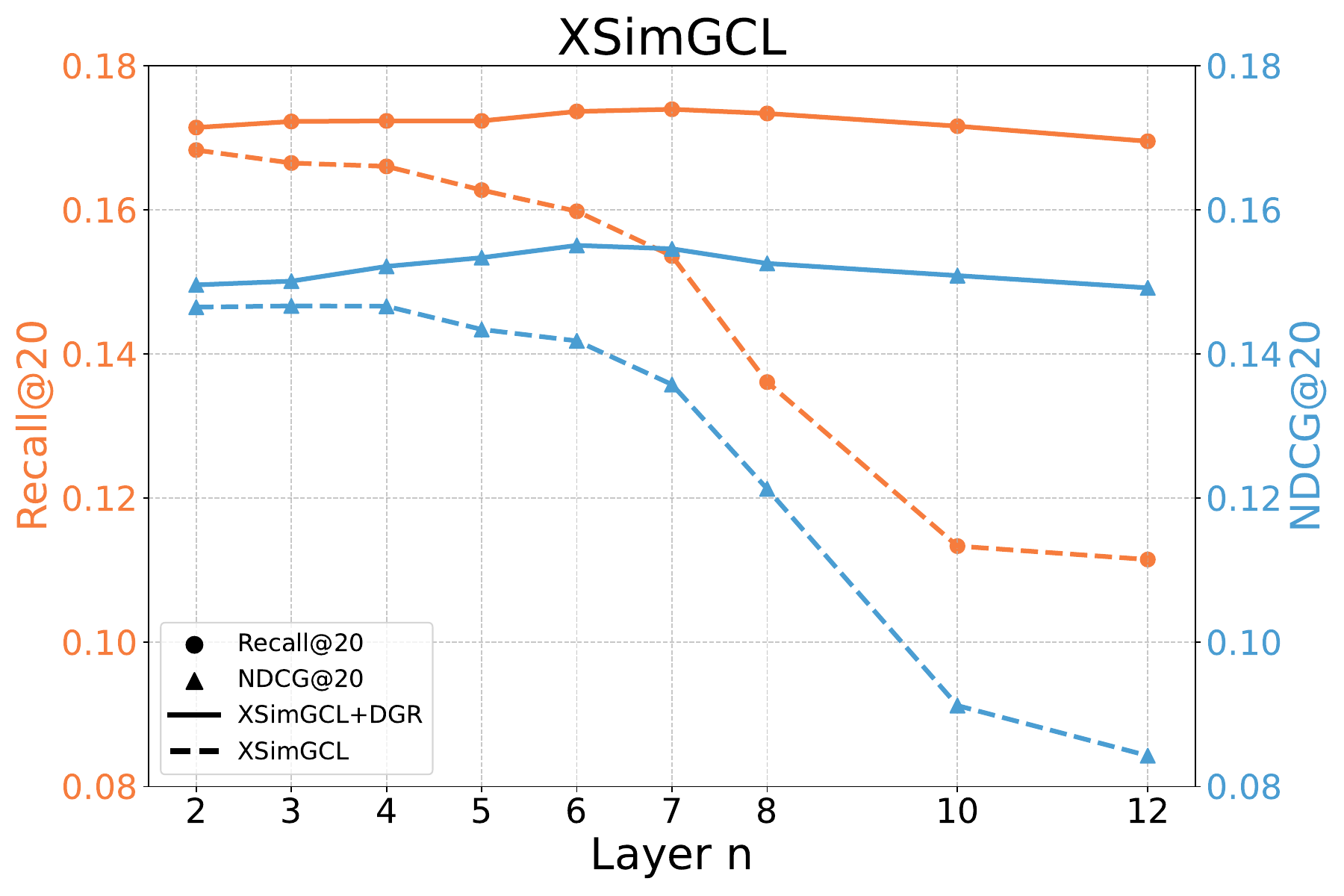}
  \includegraphics[width=0.45\textwidth]{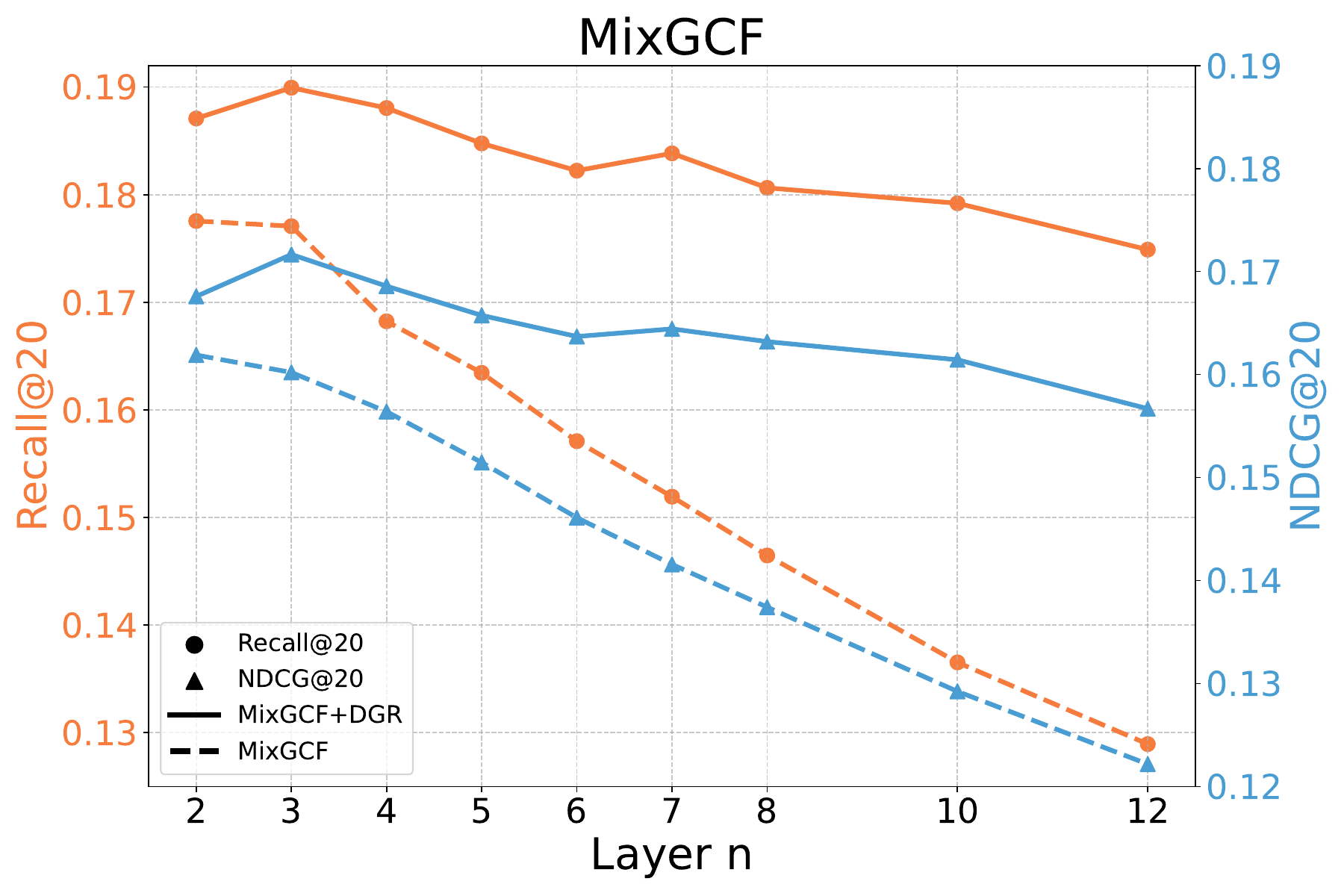}
  \vspace{-4mm}
  \caption{Performance of models with and without the DGR across a range of model depths. The orange line and blue represent Recall@20 and NDCG@20, respectively. The solid and dashed lines correspond to models with and without DGR, respectively. } 
  \vspace{-4mm}
  \label{deep_dgr_4}
\end{figure*}

The performance of baseline models with and without the DGR across a range of model depths are shown at Figure~\ref{deep_dgr_4}. The baseline models include SGL, SimGCL, XSimGCL and MixGCF.

\label{app:d}

\vspace{-2mm}
\section{Parameter Analysis of $\alpha$ }
\begin{figure}[htbp]
  \centering
    \includegraphics[width=0.327 \linewidth]{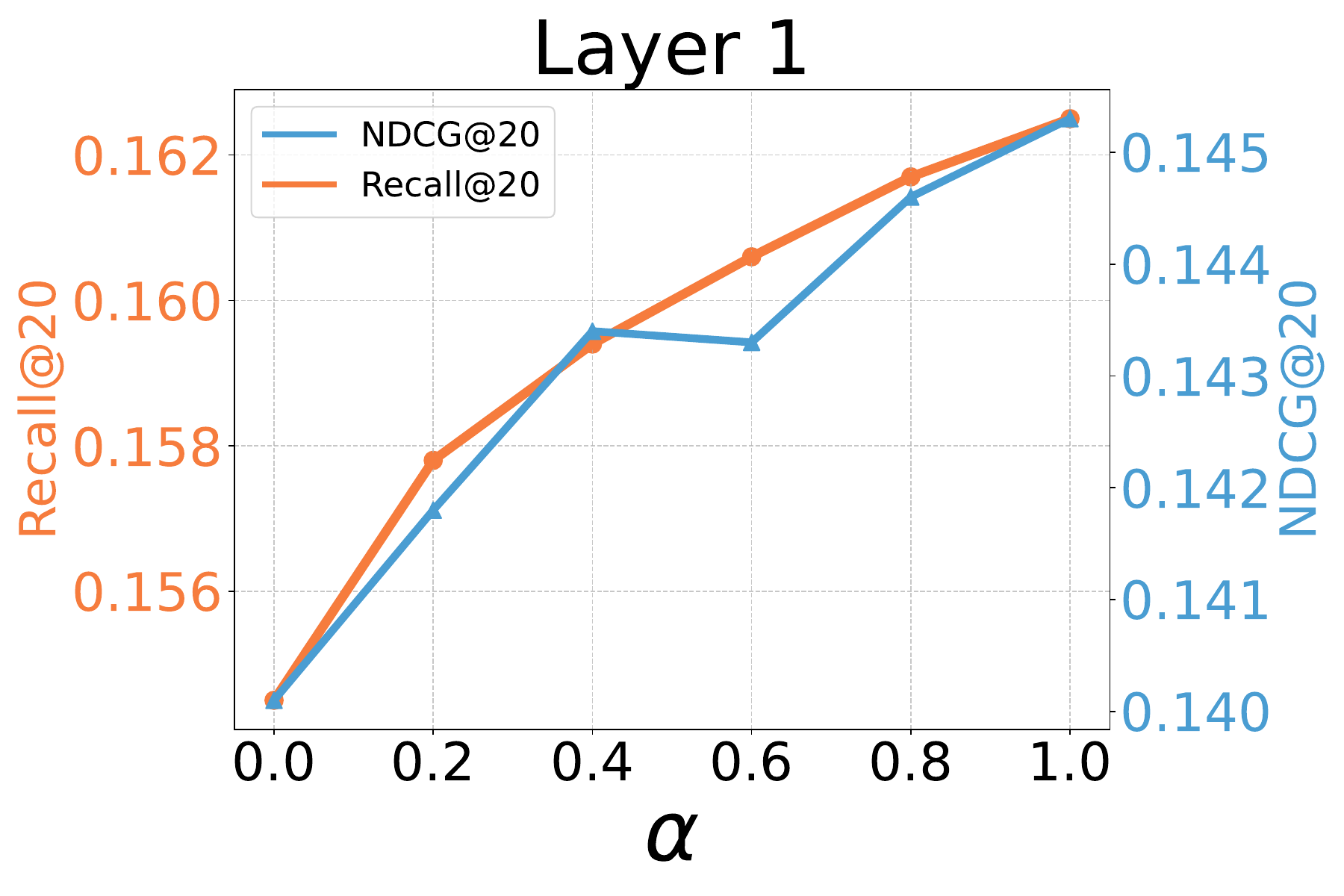}
  \hfill
    \includegraphics[width=0.327 \linewidth]{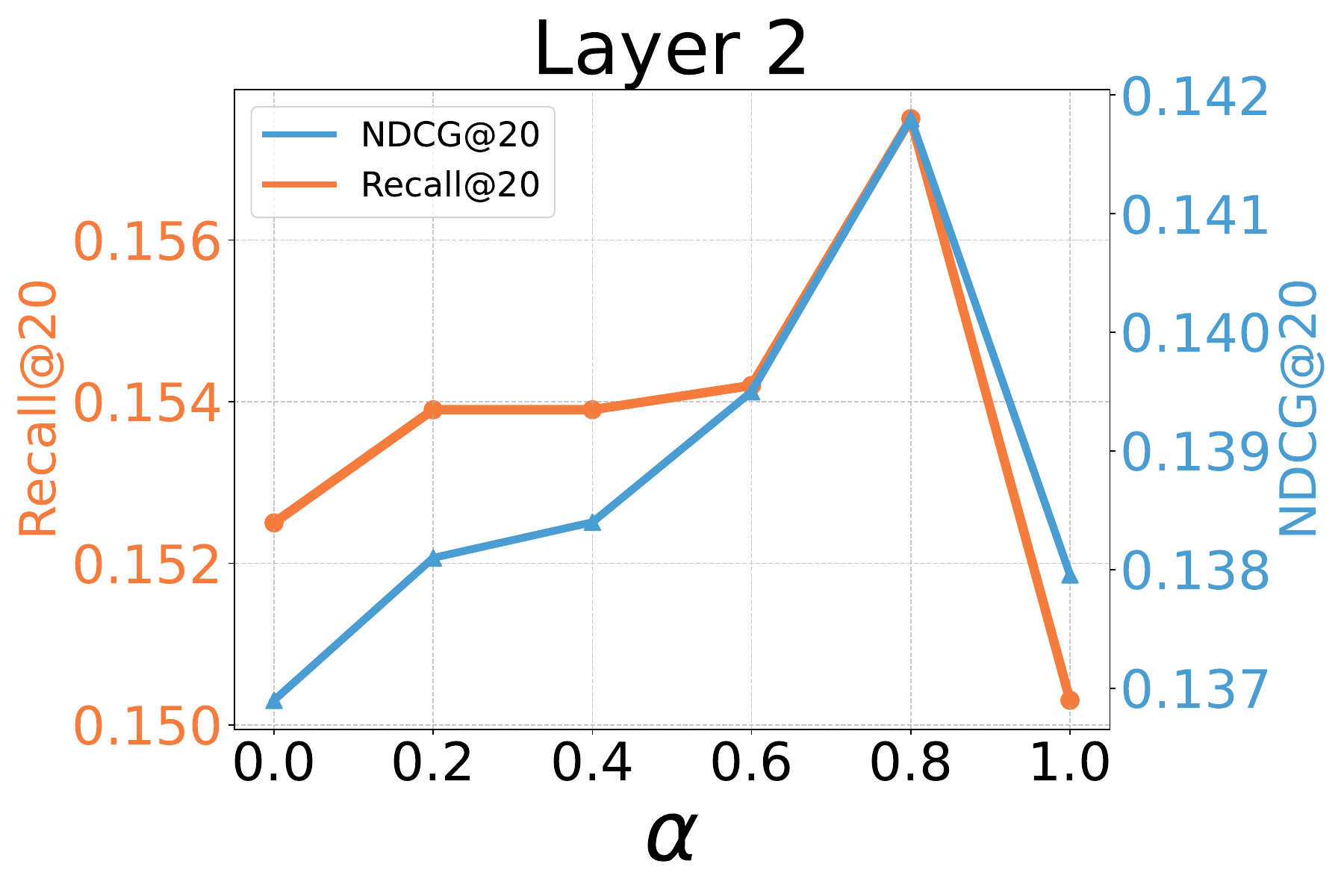}
   \hfill
      \includegraphics[width=0.327 \linewidth]{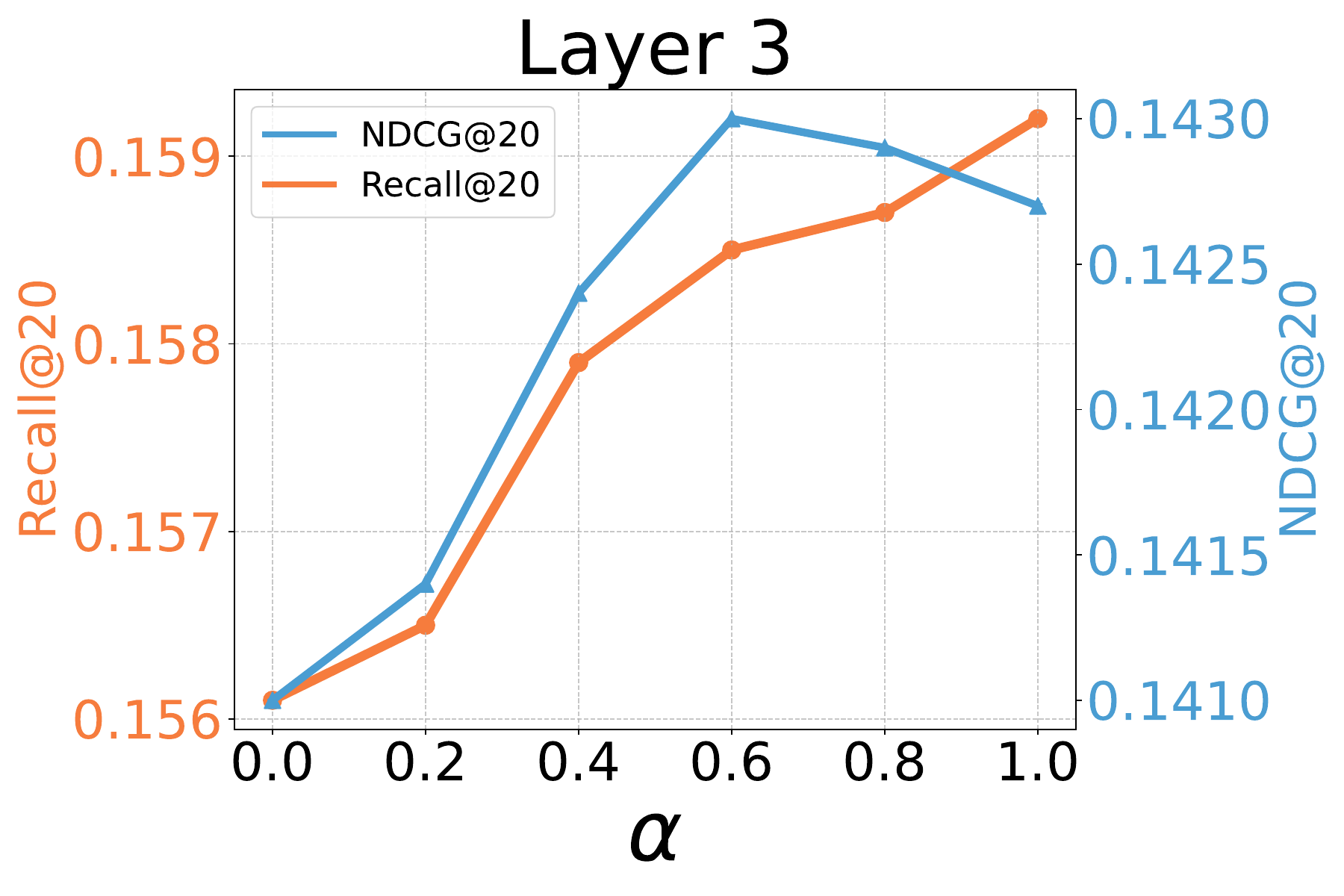}
  \caption{Parameter Analysis of $\alpha$ based on SGL+DGR.}
  \label{ove6}
\end{figure}

\begin{figure}[htbp]
  \centering
    \includegraphics[width=0.327 \linewidth]{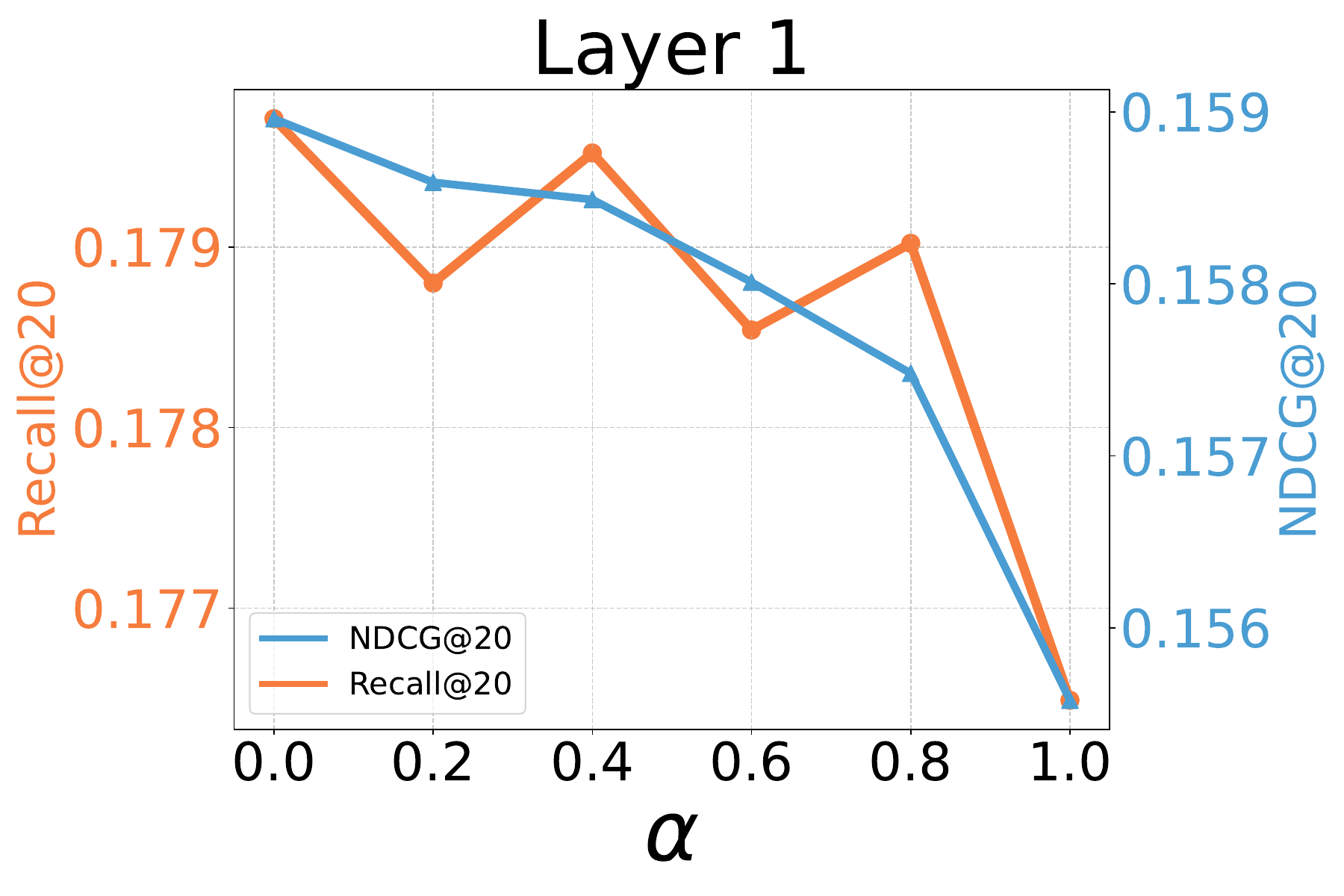}
  \hfill
    \includegraphics[width=0.327 \linewidth]{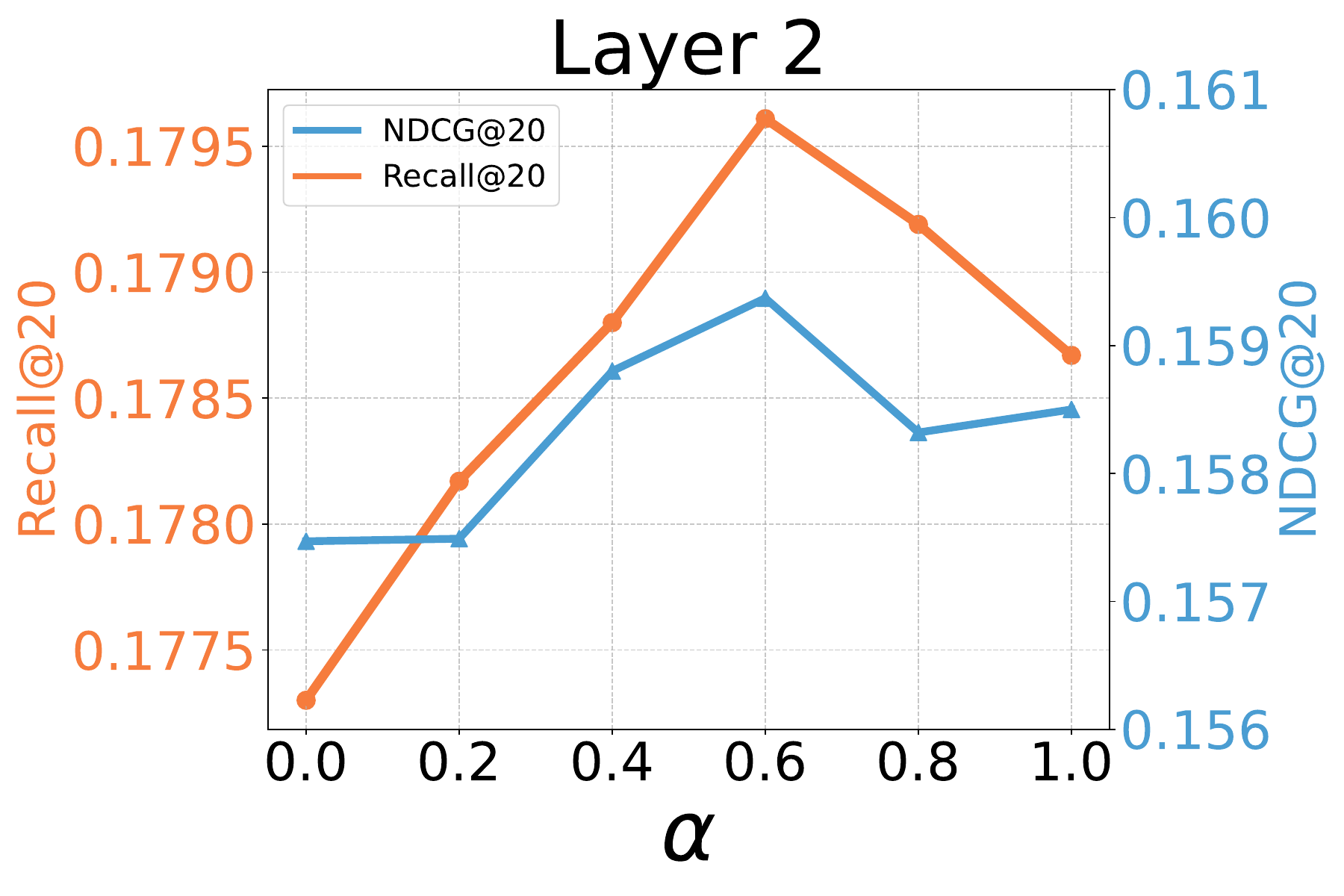}
   \hfill
      \includegraphics[width=0.327 \linewidth]{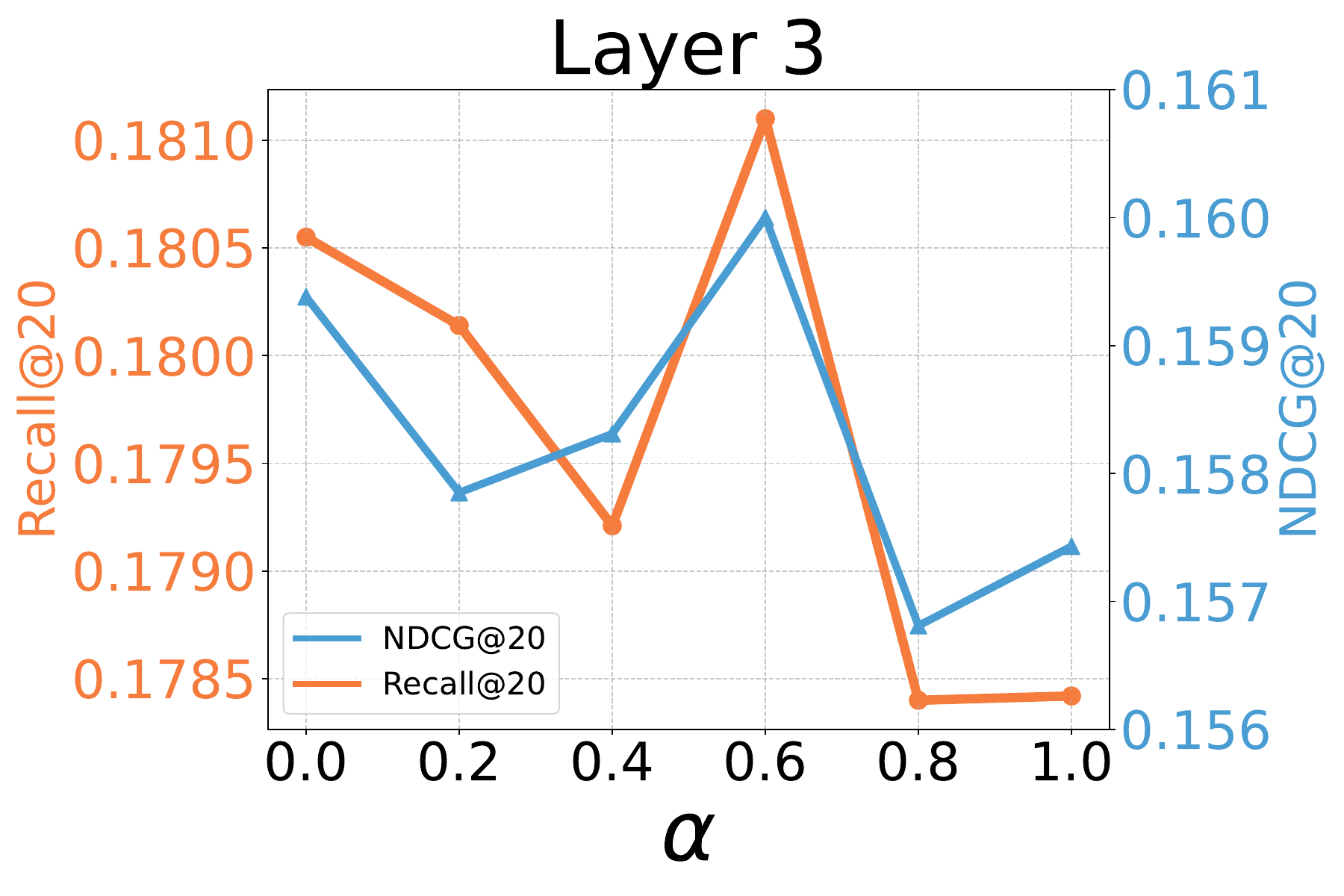}
      
  \caption{Parameter Analysis of $\alpha$ based on SimGCL+DGR.}
  \label{over7}
  
\end{figure}

\begin{figure}[htbp]
  \centering
    \includegraphics[width=0.327 \linewidth]{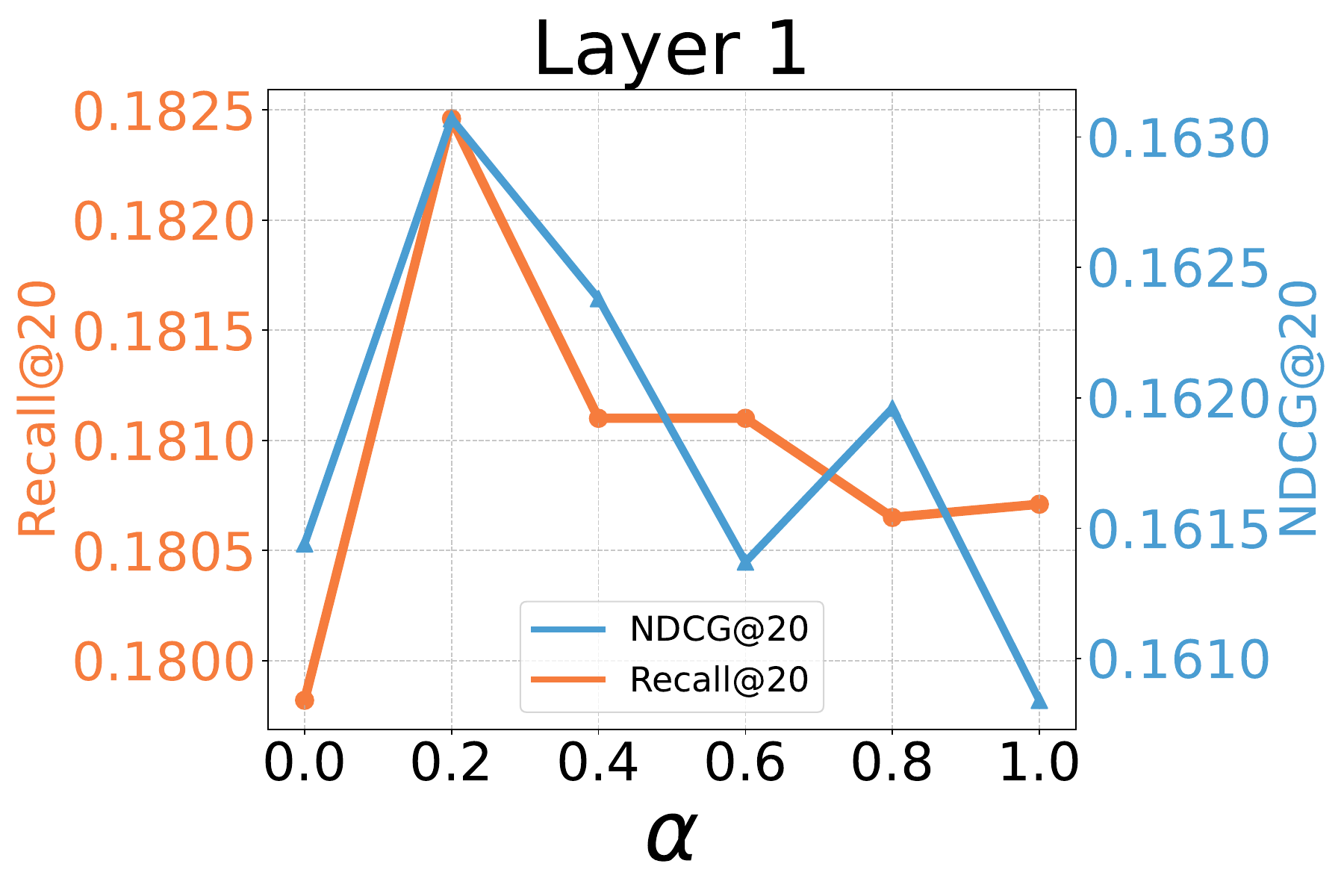}
  \hfill
    \includegraphics[width=0.327 \linewidth]{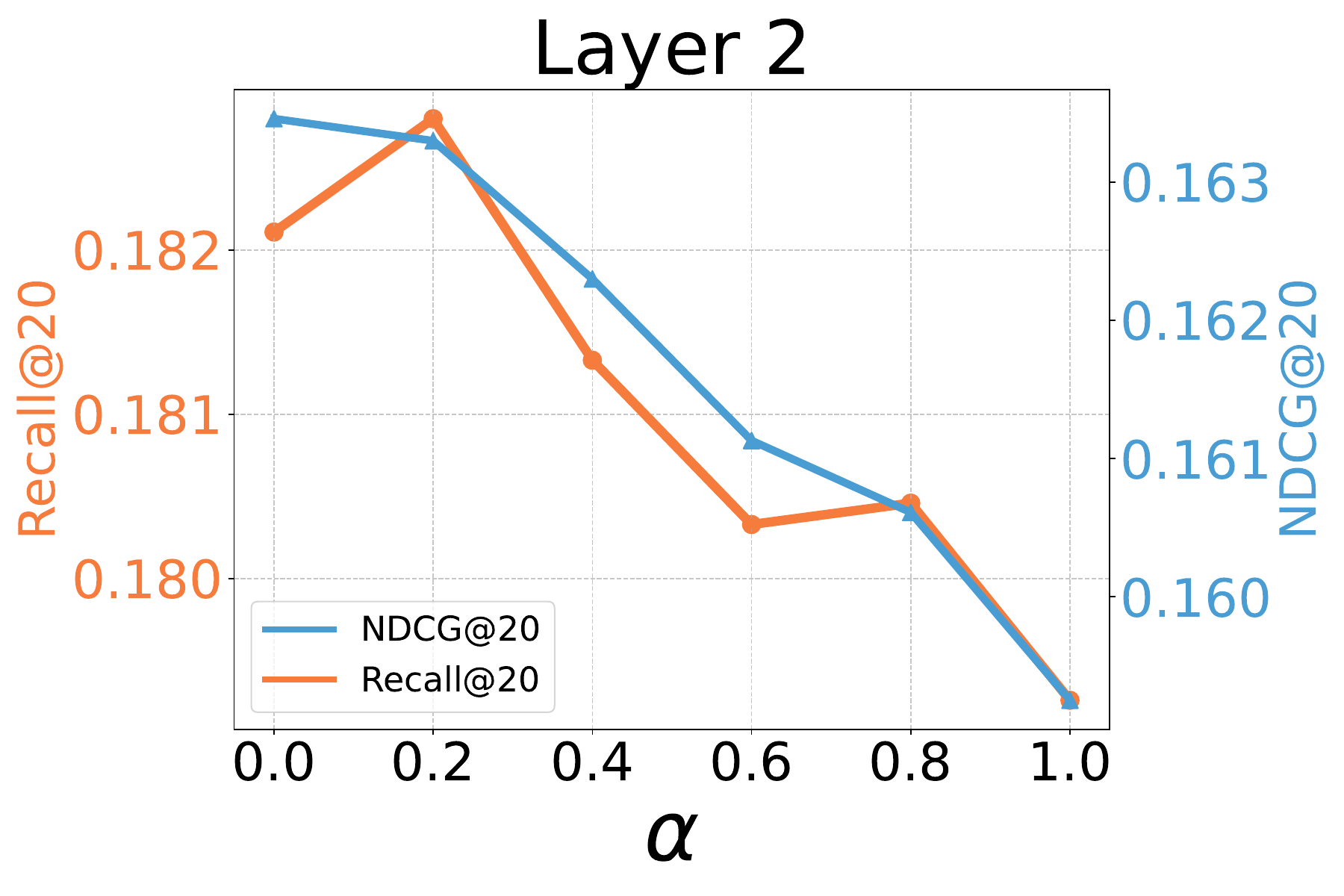}
   \hfill
      \includegraphics[width=0.327 \linewidth]{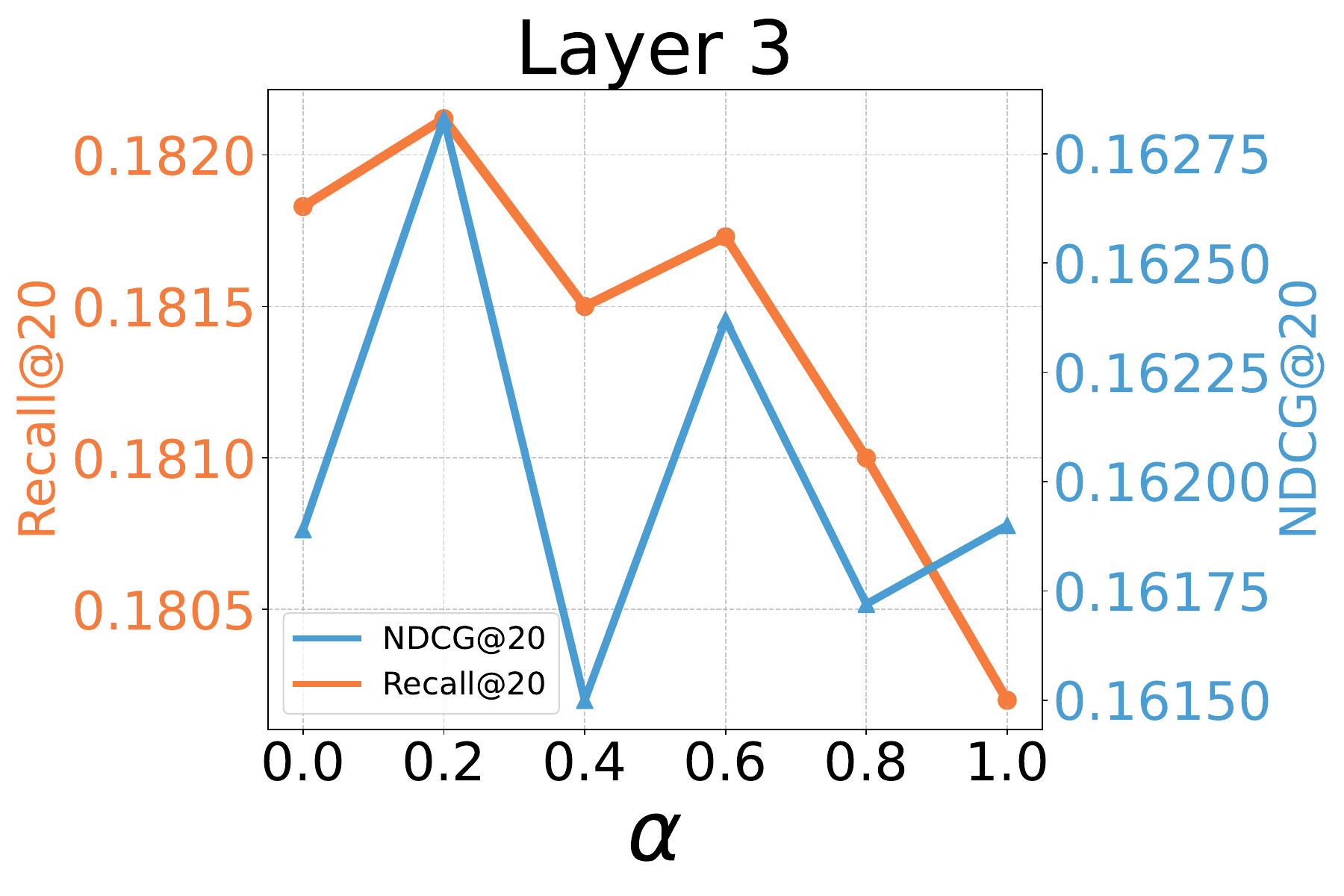}
  \caption{Parameter Analysis of $\alpha$ based on XSimGCL+DGR.}
  \label{over8}
\end{figure}

\begin{figure}[htbp]
  \centering
    \includegraphics[width=0.327 \linewidth]{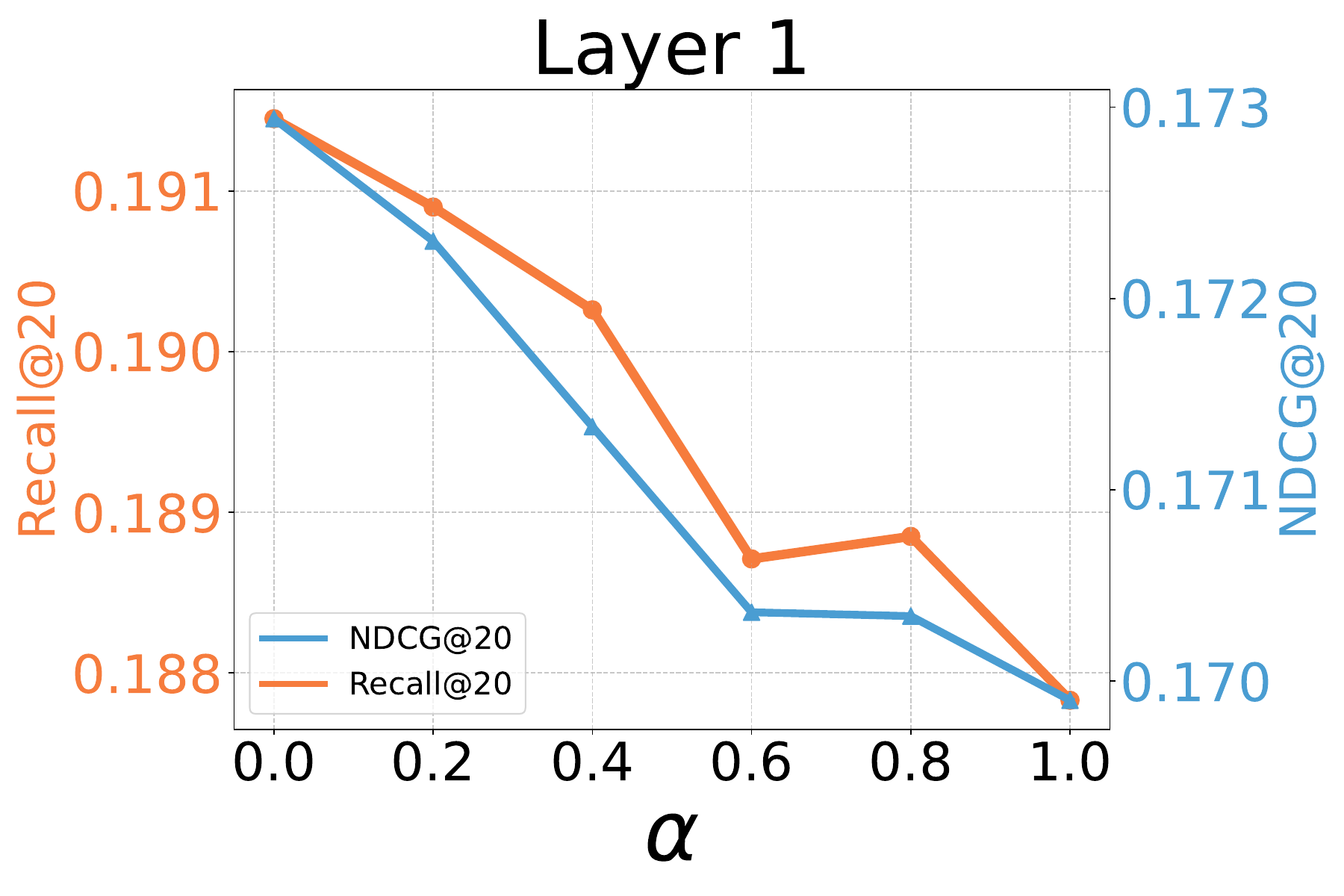}
  \hfill
    \includegraphics[width=0.327 \linewidth]{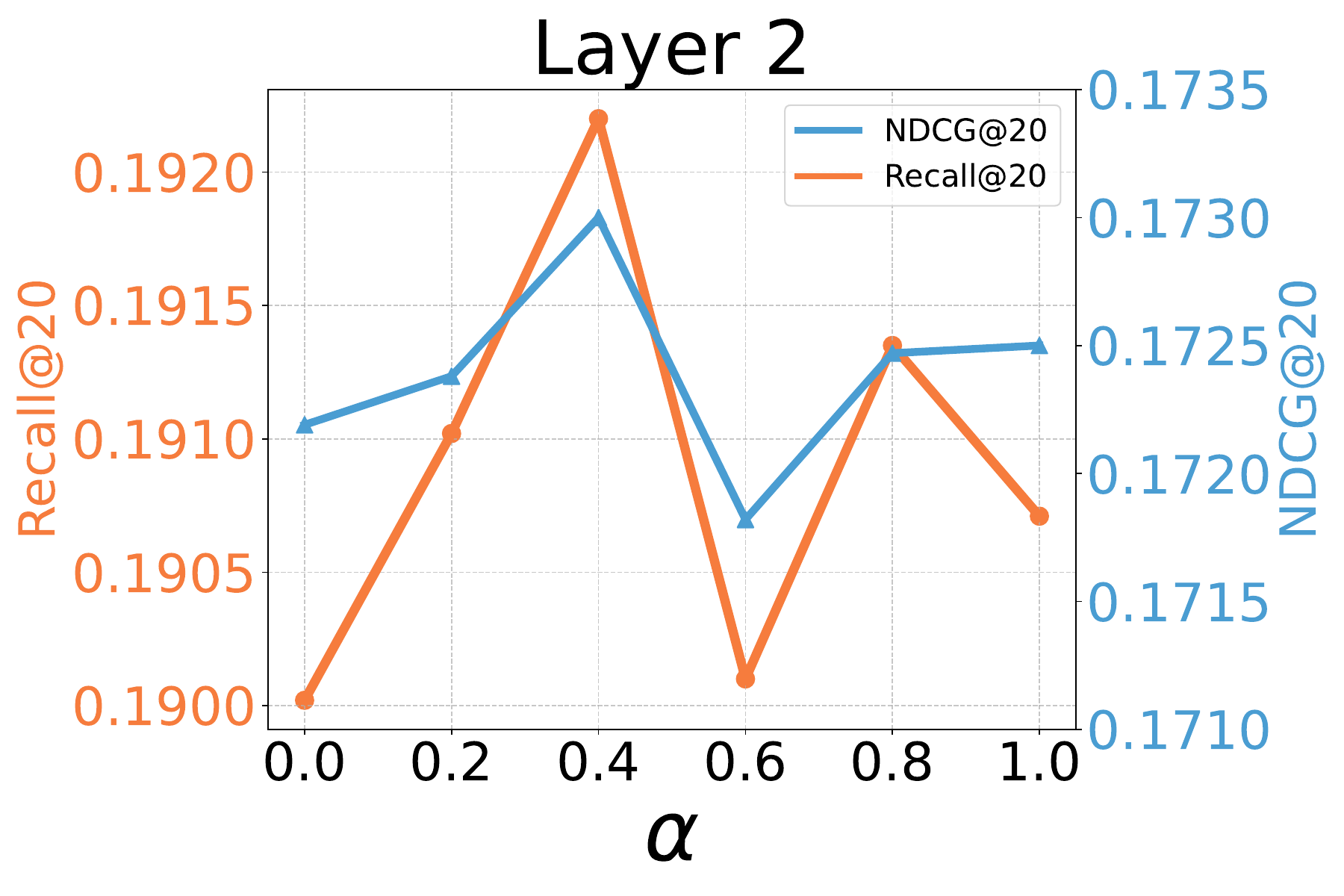}
   \hfill
      \includegraphics[width=0.327 \linewidth]{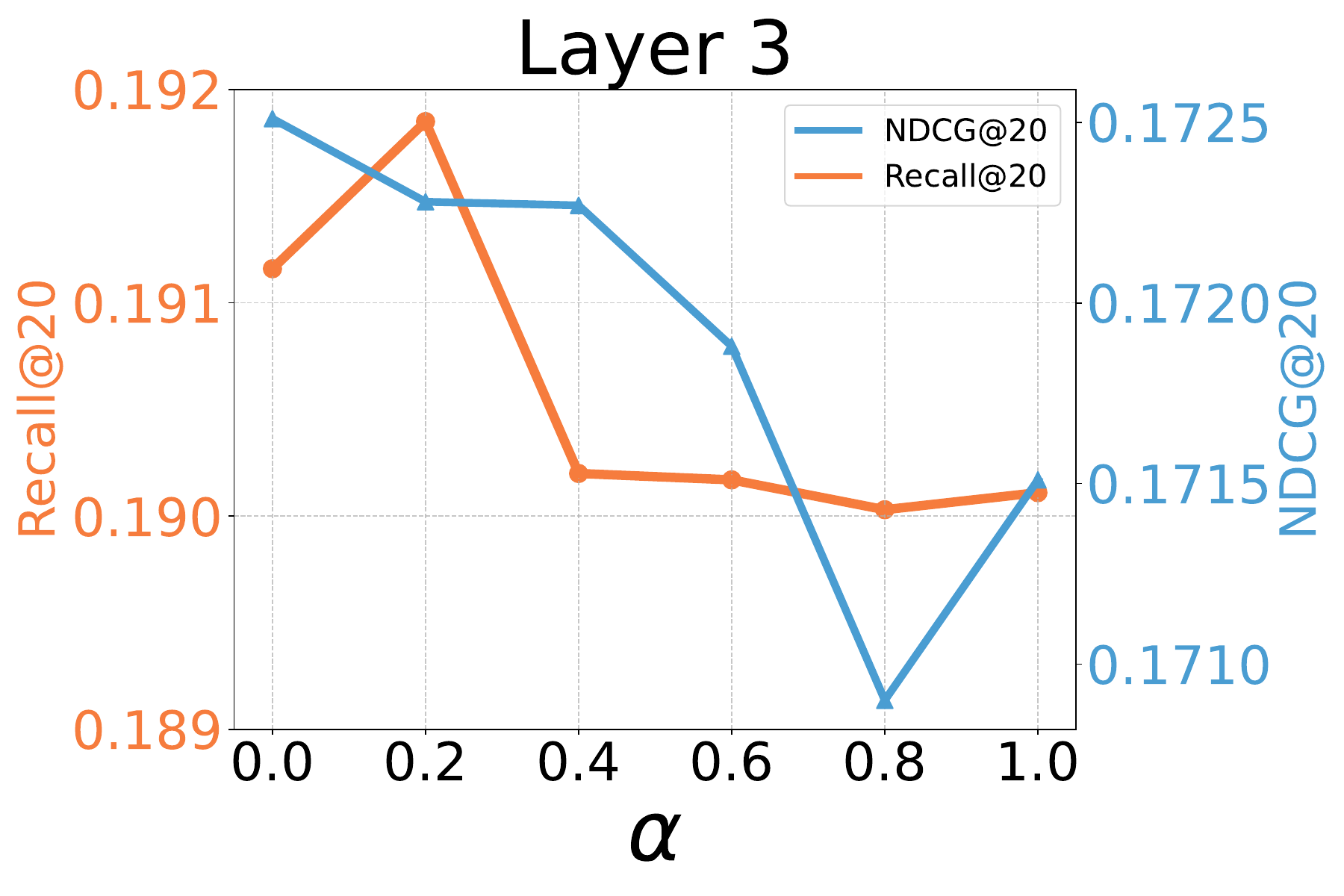}
      
  \caption{Parameter Analysis of $\alpha$ based on MixGCF+DGR.}
  \label{over9}
\end{figure}

The hyper-parameter $\alpha$ is crucial and highly sensitive to our model DGR. We conduct hyper-parameter experiments for various models and different model depths (1-3 layers). The results for SGL, SimGCL, XSimGCL and MixGCF are presented in Figure~\ref{ove6}~\ref{over7}~\ref{over8}~\ref{over9}.

\label{app:e}

\section{Parameter Analysis of $K_1$,$K_2$ and $\theta$ }
\begin{figure}[htbp]
  \centering
    \includegraphics[width=0.327 \linewidth]{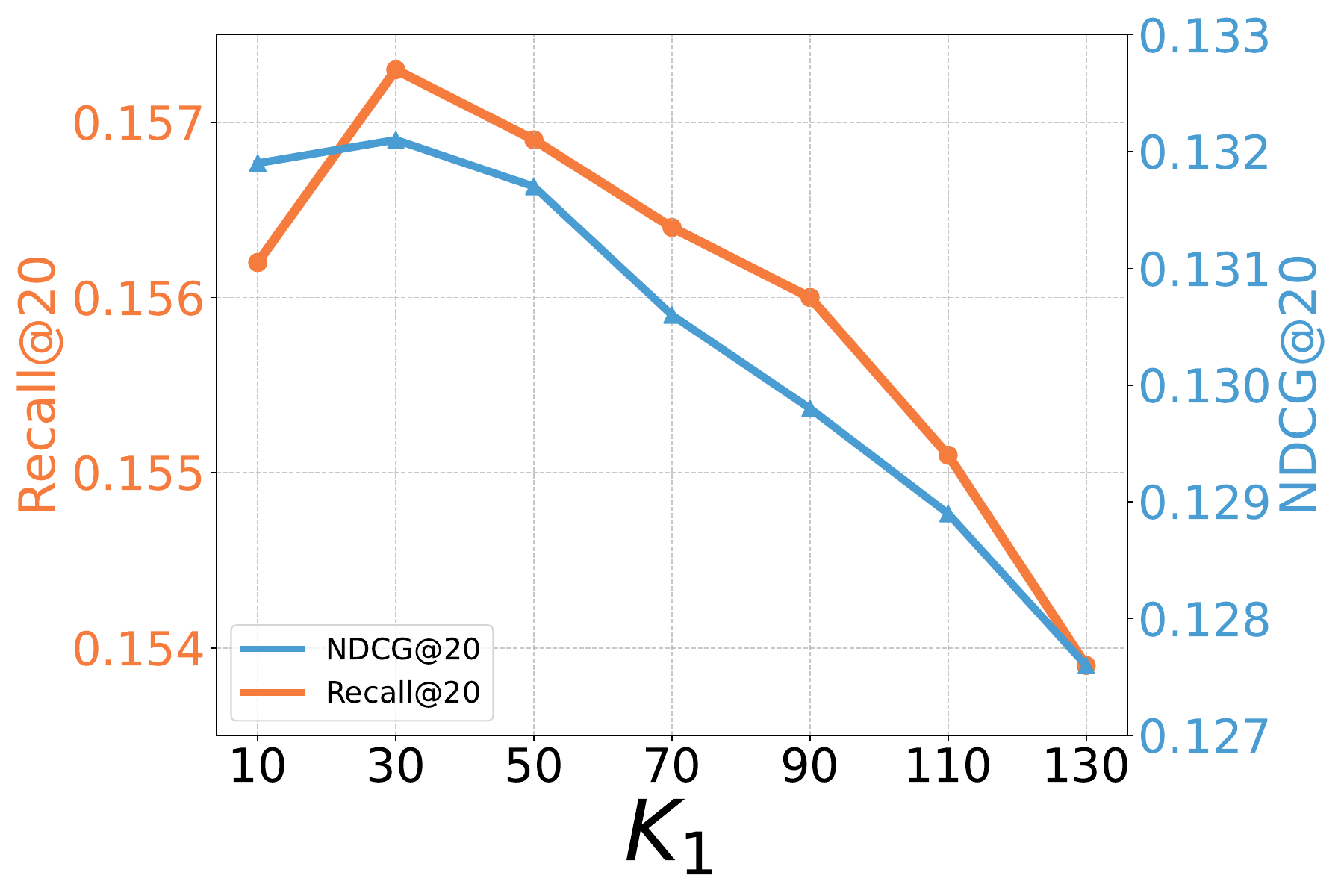}
    \hfill
    \includegraphics[width=0.327 \linewidth]{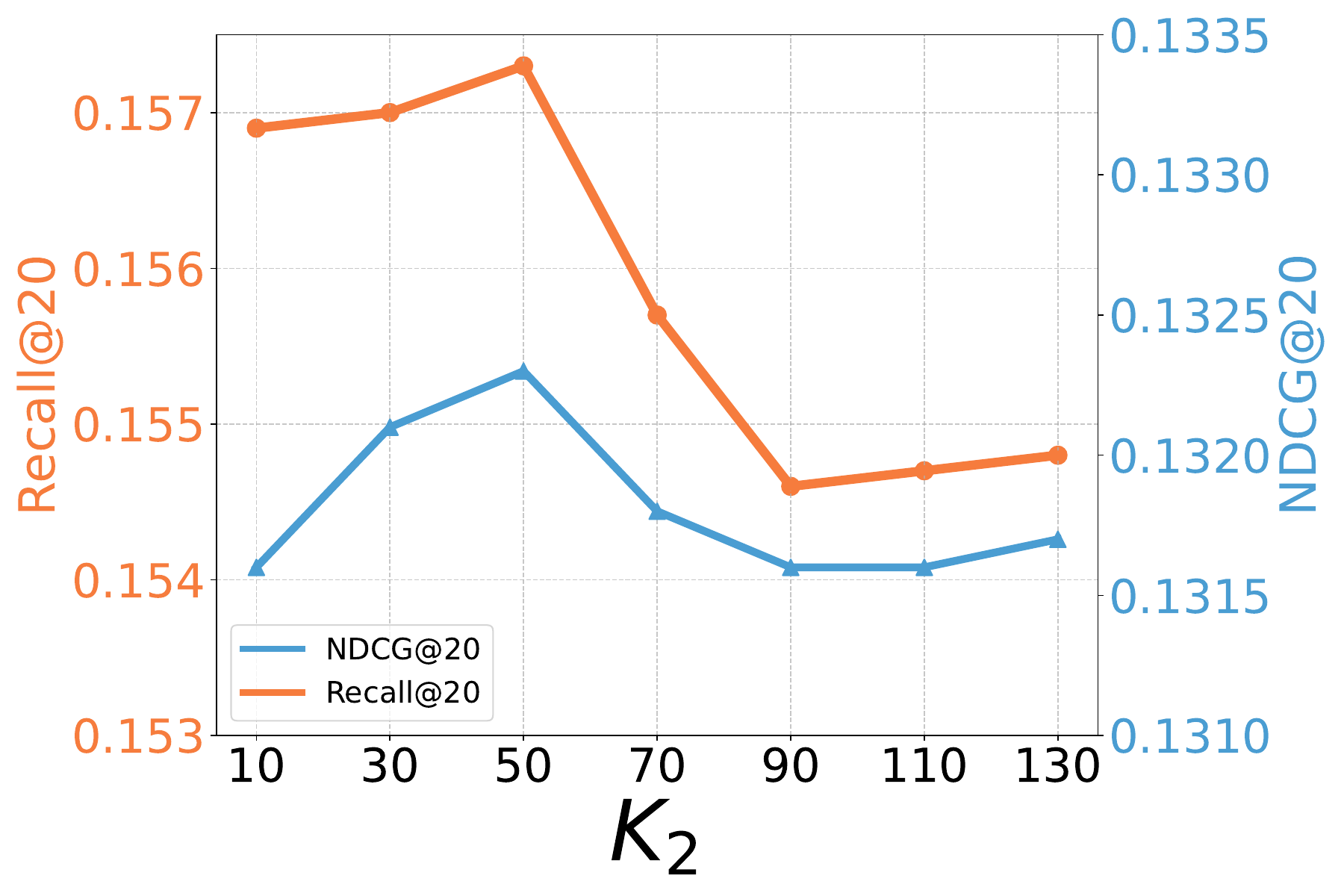}
    \hfill
    \includegraphics[width=0.327 \linewidth]{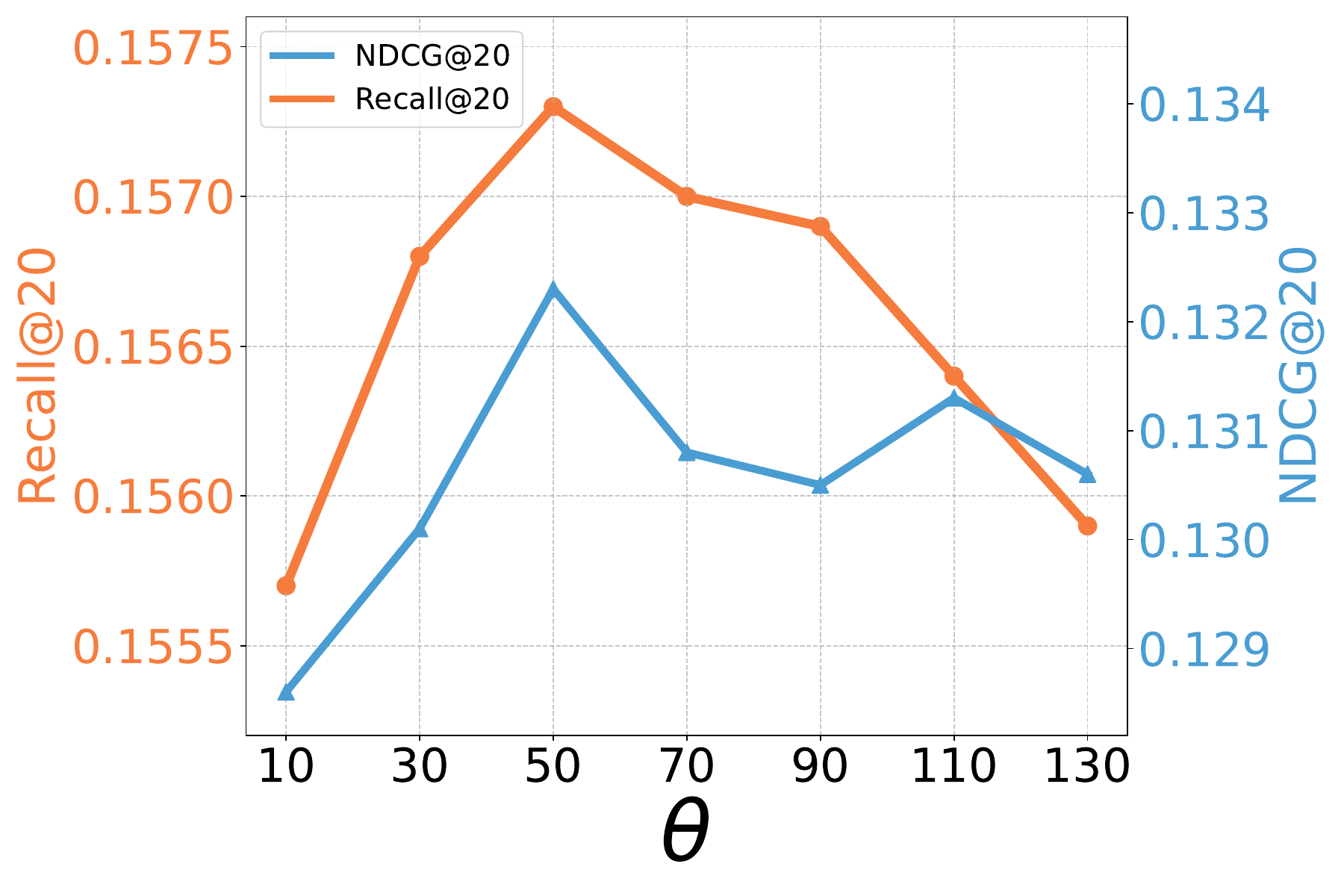}
  \caption{Parameter Analysis of $K_1$, $K_2$ and $\theta$ based on LightGCN+DGR.}
  \label{k1k2}
  \vspace{-3mm}
\end{figure}

To further investigate the impact of model results with varying values of $K_1$ $K_2$ and $\theta$, we conducted experiments using the LightGCN+DGR model on the Douban-Book dataset.
As shown in Figure~\ref{k1k2}, the best result is achieved when $K_1$ is set to 30. Similarly, the optimal outcome for $K_2$ is attained when it is set to 50 and the best result for $\theta$ is obtained when it is set to 50.

\end{document}